\documentclass[preprint,5p,authoryear]{elsarticle}

\usepackage{lineno,hyperref}
\usepackage{multicol}
\usepackage{booktabs}
\usepackage{bm}
\usepackage{xcolor}
\usepackage{threeparttable}
\usepackage{caption}
\usepackage{amssymb}
\usepackage{amsmath}
\usepackage[ruled,vlined,linesnumbered]{algorithm2e}
\SetKw{KwBy}{by}
\usepackage{paracol}
\globalcounter{algocf}

\SetCommentSty{mycommfont}
\SetNlSty{bfseries}{\color{black}}{}
\SetArgSty{textnormal}
\DontPrintSemicolon
\let\oldnl\nl
\newcommand{\nonl}{\renewcommand{\nl}{\let\nl\oldnl}}
\usepackage{physics}
\usepackage{bold-extra}

\modulolinenumbers[5]

\journal{Journal of \LaTeX\ Templates}

\hypersetup{draft}

\begin{document}

\begin{frontmatter}

\title{The MPI + CUDA Gaia AVU--GSR Parallel Solver Toward Next-generation Exascale Infrastructures}

 \author[1,2]{Valentina Cesare\corref{cor1}}
	\ead{valentina.cesare@inaf.it}
	\author[1,2]{Ugo Becciani}
	\ead{ugo.becciani@inaf.it}
	\author[2,3]{Alberto Vecchiato}
	\ead{alberto.vecchiato@inaf.it}
	\author[3]{Mario Gilberto Lattanzi}
	\ead{mario.lattanzi@inaf.it}
	\author[4]{Fabio Pitari}
	\ead{f.pitari@cineca.it}
	\author[5]{Marco Aldinucci}
	\ead{marco.aldinucci@unito.it}
	\author[3]{Beatrice Bucciarelli}
	\ead{beatrice.bucciarelli@inaf.it}
	
\cortext[cor1]{Corresponding author}
\address[1]{INAF, Astrophysical Observatory of Catania, via Santa Sofia 78, 95123 Catania, CT, Italy}
\address[2]{ICSC--Centro Nazionale di Ricerca in High Performance Computing, Big Data and Quantum Computing}
\address[3]{INAF, Astrophysical Observatory of Turin, via Osservatorio 20, 10025 Pino Torinese, TO, Italy}
\address[4]{CINECA, via Magnanelli 6/3, 40033 Casalecchio di Reno, BO, Italy}
\address[5]{University of Turin, Computer Science Department, corso Svizzera 185, 10149 Turin, TO, Italy}

\begin{abstract}

We ported to the GPU with CUDA the Astrometric Verification Unit--Global Sphere Reconstruction (AVU--GSR) Parallel Solver developed for the ESA Gaia mission, by optimizing a previous OpenACC porting of this application. The code aims to find, with a $[10,100]$~$\mu$arcsec precision, the astrometric parameters of $\sim$$10^8$ stars, the attitude and instrumental settings of the Gaia satellite, and the global parameter $\gamma$ of the parametrized Post-Newtonian formalism, by solving a system of linear equations, $\mathbf{A} \times \vb*{x} = \vb*{b}$, with the LSQR iterative algorithm. The coefficient matrix $\mathbf{A}$ of the final Gaia dataset is large, with $\sim$$10^{11}\times10^8$ elements, and sparse~\citep{Becciani_2014}, reaching a size of $\sim$10-100 TB, typical for the Big Data analysis, which requires an efficient parallelization to obtain scientific results in reasonable timescales. The speedup of the CUDA code over the original AVU--GSR solver, parallelized on the CPU with MPI + OpenMP, increases with the system size and the number of resources, reaching a maximum of $\sim$14x,  $>$9x over the OpenACC application~\citep{Cesare_OpenACC_ADASS_XXXI_in_press,Cesare_INAF_Technical_Report_OpenACC_163_2022,Cesare_2022}. This result is obtained by comparing the two codes on the CINECA cluster Marconi100, with 4 V100 GPUs per node. After verifying the agreement between the solutions of a set of systems with different sizes computed with the CUDA and the OpenMP codes and that the solutions showed the required precision, the CUDA code was put in production on Marconi100, essential for an optimal AVU--GSR pipeline and the successive Gaia Data Releases. This analysis represents a first step to understand the (pre-)Exascale behavior of a class of applications that follow the same structure of this code. In the next months, we plan to run this code on the pre-Exascale platform Leonardo of CINECA, with 4 next-generation A200 GPUs per node, toward a porting on this infrastructure, where we expect to obtain even higher performances. 

\end{abstract}


\end{frontmatter}

\section{Introduction}
\label{sec:Intro}

In this epoch of technological evolution, the size of the problems to solve in several contexts is rapidly increasing and can also require up to $\sim$10-100 PB of storage. To allow the analysis of these \textit{Big Data}, novel parallelization techniques have to be continuously defined to find solutions in human-size timescales. The architecture of the infrastructures is also consequently changing, becoming increasingly heterogeneous~\citep{Carpenter_2022}, to accomplish the necessity of optimally computing data of these sizes, going toward the (pre-)Exascale era. The supercomputers will require an increasing number of computational nodes, which will have, in turn, a RAM memory organized in a multi-levels hierarchy of nonvolatile memories and hosts less performant than the accelerators (such as GPUs or FPGAs) that will be increasingly employed for calculations and will have an increasing memory and number of streaming multiprocessors. This configuration will also be likely to achieve the target of Green Computing, namely to process this amount of Big Data without excessively increasing the energy consumption while obtaining a high performance. Moreover, the infrastructures will need increasingly faster bridges between the CPU and the accelerators, to reduce the host-to-device (H2D) and the device-to-host (D2H) data transfers bottleneck in the applications, and storage areas defined with parallel filesystems to guarantee a faster access to the data~\citep{Carpenter_2022}. Computer clusters such as Marconi100 (M100) of CINECA\footnote{\url{https://www.hpc.cineca.it/hardware/marconi100}} and JUWELS of Forschungszentrum J{\"u}lich\footnote{\url{https://fz-juelich.de/ias/jsc/EN/Expertise/Supercomputers/JUWELS/JUWELS_node.html}} are already going in this direction but a turning point will be provided by next-generation pre-Exascale infrastructures, such as the CINECA platform Leonardo\footnote{\url{https://www.cineca.it/temi-caldi/Leonardo}}, which started to be operative last November.
  
A typical science case that might involve a large amount of data is the \textit{inverse problem}, that consists in estimating the parameters of a model from a set of observational measurements. Two possible approaches for this task are the \textit{frequentist} and the \textit{bayesian} ones. Concerning the former approach, one of the exploited computational techniques is the \textit{LSQR} iterative algorithm, to solve large, ill-posed, overdetermined, and possibly sparse systems of equations~\citep{Paige_and_Saunders_1982a,Paige_and_Saunders_1982b}. This algorithm is employed in several contexts, such as medicine~\citep{Bin_2020,Guo_2021}, geophysics~\citep{Joulidehsar_2018,LIANG_2019,LSQR_geology_2019}, geodesy~\citep{Baur_and_Austen_2005}, industry~\citep{Jaffri_2020}, and astronomy~\citep{Borriello_1986,VanderMarel_1988,Naghibzadeh_and_vanderVeen_2017,Becciani_2014,Cesare_OpenACC_ADASS_XXXI_in_press,Cesare_INAF_Technical_Report_OpenACC_163_2022,Cesare_INAF_Technical_Report_CUDA_164_2022,Cesare_2022}. For a more in-depth discussion about the LSQR algorithm and other LSQR-based applications and libraries, see Section 2 of~\citet{Cesare_2022}. 

As an example, in the astronomy context, this algorithm is employed by the Gaia Astrometric Verification Unit--Global Sphere Reconstruction (AVU–-GSR) Parallel Solver. This code was developed for the ESA Gaia mission~\citep{Gaia_Collaboration_DR3_Vallenari_2022} under the Data Processing and Analysis Consortium (DPAC)~\citep{Mignard_and_Drimmel_2007}, i.e., the scientific community of the mission, funded by the national space agencies, in charge of the definition of the data reduction pipelines~\citep{Vecchiato_2018}. The code has been in production since 2014 on M100 cluster according to an agreement between Istituto Nazionale di Astrofisica (INAF) and CINECA, with the support of the Italian Space Agency (ASI). 

The Gaia AVU–-GSR code solves with the LSQR algorithm an overdetermined system of linear equations~\citep{Becciani_2014,Cesare_2022},
\begin{equation}
\label{eq:Axb}
\mathbf{A} \times \vb*{x} = \vb*{b},
\end{equation}
 where $\mathbf{A}$ is the coefficient matrix, and $\vb*{b}$ and $\vb*{x}$ are the arrays of the known terms and of the solution, respectively. The matrix $\mathbf{A}$ is sparse and it might contain $\sim$$10^{11} \times 10^8$ elements for the expected final dataset of Gaia. Even only considering its non-zero coefficients, it will occupy a large amount of memory ($\sim$10-100~TB). 
	
By solving this system, the AVU-GSR code finds, with an accuracy in the range of 10-100~$\mu$arcsec and of 10-100~$\mu$arcsec~year$^{-1}$, the astrometric parameters (parallaxes, right ascension, declination, and proper motions along these two directions) of $\sim$$10^8$ stars in the Milky Way, the so-called primary stars~\citep{Vecchiato_2018}. 
The Gaia AVU–-GSR code is a verification module of the same solution found with the software Astrometric Global Iterative Solution (AGIS;~\citealt{OMullane_2011,Lindegren_2012}) adopting a different algorithm, to make the determination of the astrometric parameters more robust.
Besides the astrometric parameters, the Gaia AVU–-GSR solver finds the attitude and instrumental specifications of the Gaia satellite, and the global parameter $\gamma$ of the Parametrized Post-Newtonian (PPN) formalism, with the same precision around $[10,100]$~$\mu$arcsec. The high accuracy of these parameters is essential to properly investigate the formation and the evolution of the Milky Way~\citep[e.g.,][]{Giammaria_2021,Krolikowski_2021} and to test Einstein's theory of General Relativity~\citep[e.g.,][]{Vecchiato_2003,Hees_2018,Crosta_2020,Butkevich_2022}. 

The LSQR algorithm is the bulk of the AVU--GSR solver and it works by calculating, at each iteration, the iterative estimates of the known terms and of the solution arrays with the \textit{aprod 1} and \textit{aprod 2} functions:
\begin{equation}
\label{eq:aprod1}
\vb*{b}^i += \mathbf{A} \times \vb*{x}^{i-1},
\end{equation}
and
\begin{equation}
\label{eq:aprod2}
\vb*{x}^i += \mathbf{A}^T \times \vb*{b}^i,
\end{equation}
which are the most computational demanding parts of the LSQR procedure, representing more than 90\% of the entire calculation.

The last official in-production version of the Gaia AVU--GSR code was entirely parallelized on the CPU with a hybrid MPI + OpenMP approach. In~\citet{Cesare_2022}, we explored the feasibility of a GPU porting of the application by adopting a preliminary approach, where we replaced the OpenMP directives with the OpenACC ones. With this porting, the speedup of the OpenACC code over the OpenMP code, both run on M100, was of $\sim$1.5. In this paper, we present an optimization of the GPU parallelization of the code starting from the results of our first porting, where we replace the high-level parallelization approach, using OpenACC, with a low-level one, using CUDA~\citep{Cesare_INAF_Technical_Report_CUDA_164_2022}. This implied a reorganization of several parts of the code but a substantial performance boost of $\sim$$14$x over the OpenMP version, as tested on M100. This speedup might further improve on Leonardo, with GPUs having a larger memory and number of streaming multiprocessors than on M100, which is an optimistic estimate in perspective of a future porting on this platform. The CUDA code also showed to achieve a great numerical stability and to obtain parameters with the required accuracy, reasons for which it was put in production on M100.

The paper develops across the following sections. Section~\ref{sec:Gaia_AVU_GSR_code_structure} summarizes the general structure of the Gaia AVU--GSR code, Section~\ref{sec:Gaia_AVU_GSR_code_structure_OpenMP_OpenACC} describes the previous versions of the code, i.e., the MPI + OpenMP (Section~\ref{sec:Gaia_AVU_GSR_code_structure_OpenMP}) and the MPI + OpenACC (Section~\ref{sec:Gaia_AVU_GSR_code_structure_OpenACC}) ones, and Section~\ref{sec:Gaia_AVU_GSR_code_CUDA_porting} details the CUDA porting of the Gaia AVU--GSR code. A performance comparison with the OpenACC porting is presented throughout Section~\ref{sec:Gaia_AVU_GSR_code_CUDA_porting}. Section~\ref{sec:Gaia_AVU_GSR_code_performance_tests} compares the performance of the MPI + CUDA and the MPI + OpenMP codes on M100 for a set of systems with increasing size and Section~\ref{sec:Gaia_AVU_GSR_code_Numerical_stability} compares the solutions of these systems to verify their consistency and quantify the accuracy of the obtained solutions. At last, Section~\ref{sec:Conclusions_and_future_works} discuss the main results of the paper and presents the future analyses to be developed.

\section{The structure of the Gaia AVU--GSR code}
\label{sec:Gaia_AVU_GSR_code_structure}

The black part of Algorithm~\ref{algo:Full_application} summarizes the general structure of the Gaia AVU--GSR code, which is common to the OpenMP, OpenACC, and CUDA versions. The preparatory phase consists in importing from binary files the quantities necessary to solve the system (e.g., the coefficient matrix and the known terms; line~\ref{algo:Full_application:Data_import}). To accelerate the convergence speed of the iterative procedure, the system is preconditioned before the starting of the LSQR algorithm. Specifically, we normalized the parameters of each column by the norm of the column itself (lines~\ref{algo:Full_application:precond_1}--\ref{algo:Full_application:precond_2}). The normalization factors of all the columns are stored in a 1D array, $\vb*{p}$. The solution is re-multiplied by $\vb*{p}$ after the end of the LSQR algorithm (line~\ref{algo:Full_application:precond_3}).  Then, the initial guess of the solution to be iteratively found with the LSQR algorithm is computed through the \textit{aprod 2} function (see Eq.~\eqref{eq:aprod2}; line~\ref{algo:Full_application:aprod2_IS}). Each MPI process calculates a part of the solution that is then reduced among all the MPI processes (line~\ref{algo:Full_application:MPI_reduce_x_aprod2_IS}).

After these passages, the LSQR procedure starts. The LSQR algorithm is a while loop (lines~\ref{algo:Full_application:LSQR_start}--\ref{algo:Full_application:LSQR_end}) that iterates the solution up to a convergence condition or until a maximum number of iterations set at runtime is reached. At each iteration, the two main steps are the execution of the \textit{aprod} function in the modes 1 and 2 (lines~\ref{algo:Full_application:aprod1} and~\ref{algo:Full_application:aprod2}). The \textit{aprod 1} (Eq.~\eqref{eq:aprod1}) provides the iterative estimate of the known terms $\vb*{b}$ for each equation of the system and for a set of constraints equations~\citep{Vecchiato_2018}, required since the system is overdetermined. After the calculation of the \textit{aprod 1}, the $\vb*{b}$ array is reduced among the MPI processes. Then, the \textit{aprod 2} (Eq.~\eqref{eq:aprod2}) provides the iterative estimate of the solution array $\vb*{x}$. Also for this step, the constraints equations are defined and the solution is reduced among the MPI processes. At the end of each iteration, the errors (variances) on the unknowns and the covariances between the different couples of unknowns are calculated (line~\ref{algo:Full_application:var_covar}). The convergence condition is achieved in the least-squares sense, when the residuals $r^i = b^i - A \times x^i$, estimated at the $i$-th iteration, go below a given tolerance, set to the machine precision ($\sim$$10^{-16}$ on M100).
 
\begin{algorithm*}
	Import data (e. g., $\mathbf{A}$, $\vb*{b}$) from files \; \label{algo:Full_application:Data_import}
	Calculate preconditioning array $\vb*{p}$ \; \label{algo:Full_application:precond_1}
	Normalization of $\mathbf{A}$ by $\vb*{p}$ \; \label{algo:Full_application:precond_2}
	\tcp{Get the number of the devices in the node}
	\textcolor{gray}{\texttt{\textbf{cuda\_error = cudaGetDeviceCount(\&deviceCount)}} \label{algo:Full_application:CUDA_get_number_GPUs_in_node}} \;
	\tcp{Set the number of the device in the node}
	\textcolor{gray}{\texttt{\textbf{cuda\_error = cudaSetDevice(pid \% deviceCount)}} \label{algo:Full_application:CUDA_set_number_GPU_in_node}} \;
	\textcolor{gray}{\texttt{\textbf{cuda\_error = cudaMalloc}}($\&\mathbf{A}_\textbf{d,dev}$,length($\mathbf{A}_\textbf{d}$)) \label{algo:Full_application:cudaMalloc_Example}} \;
	\textcolor{gray}{\textbf{...}} \;
	\textcolor{gray}{\texttt{\textbf{cuda\_error = cudaMemcpy}}($\mathbf{A}_\textbf{d,dev}$,$\mathbf{A}_\textbf{d}$,length($\mathbf{A}_\textbf{d}$),\texttt{\textbf{cudaMemcpyHostToDevice}})} \; \label{algo:Full_application:copyin_before_LSQR_beg}
	\textcolor{gray}{\texttt{\textbf{cuda\_error = cudaMemcpy}}($\mathbf{M}_\textbf{i,dev}$,$\mathbf{M}_\textbf{i}$,length($\mathbf{M}_\textbf{i}$),\texttt{\textbf{cudaMemcpyHostToDevice}})} \;
	\textcolor{gray}{\texttt{\textbf{cuda\_error = cudaMemcpy}}($\mathbf{I}_\textbf{c,dev}$,$\mathbf{I}_\textbf{c}$,length($\mathbf{I}_\textbf{c}$),\texttt{\textbf{cudaMemcpyHostToDevice}})} \;
	\textcolor{gray}{\texttt{\textbf{cuda\_error = cudaMemcpy}}($\vb*{b}_{\rm dev}$,$\vb*{b}$,length($\vb*{b}$),\texttt{\textbf{cudaMemcpyHostToDevice}})} \; \label{algo:Full_application:copyin_b_before_LSQR}
	\textcolor{gray}{\texttt{\textbf{cuda\_error = cudaMemcpy}}($\vb*{x}_{\rm dev}$,$\vb*{x}$,length($\vb*{x}$),\texttt{\textbf{cudaMemcpyHostToDevice}})} \; \label{algo:Full_application:copyin_before_LSQR_end}
	\tcp{Calculation of the initial solution $\vb*{x}_0$}
	\textit{aprod} 2 call (in each MPI process) \; \label{algo:Full_application:aprod2_IS}
	\textcolor{gray}{\texttt{\textbf{cuda\_error = cudaMemcpy}}($\vb*{x}$,$\vb*{x}_{\rm dev}$,length($\vb*{x}$),\texttt{\bf cudaMemcpyDeviceToHost})} \; \label{algo:Full_application:copyout_x_aprod2_IS}
	\tcp{Reduction of the initial solution among the MPI processes}
	\texttt{MPI\_Allreduce}($\vb*{x}_0$)\; \label{algo:Full_application:MPI_reduce_x_aprod2_IS}
	\textcolor{gray}{\texttt{\textbf{cuda\_error = cudaMemcpy}}($\vb*{x}_{\rm dev}$,$\vb*{x}$,length($\vb*{x}$),\texttt{\textbf{cudaMemcpyHostToDevice}})} \; \label{algo:Full_application:copyin_x_aprod2_IS}
	\tcp{LSQR algorithm}
	\While{(conv. cond. $\vert$ $\vert$ max itn. reached)}{ \label{algo:Full_application:LSQR_start}
		\tcp{Iterative estimate of the known terms array $\vb*{b}$}
		\textit{aprod} 1 call (in each MPI process)\; \label{algo:Full_application:aprod1}
		\textcolor{gray}{\texttt{\textbf{cuda\_error = cudaMemcpy}}($\vb*{b}$,$\vb*{b}_{\rm dev}$,length($\vb*{b}_{\rm Constraints}$),\texttt{\textbf{cudaMemcpyDeviceToHost}})} \; \label{algo:Full_application:copyout_b_aprod1}
		\tcp{Reduction of $\vb*{b}$ among the MPI processes}
		\texttt{MPI\_Allreduce($\vb*{b}$)}\; \label{algo:Full_application:MPI_reduce_b_aprod1}
		\textcolor{gray}{\texttt{\textbf{cuda\_error = cudaMemcpy}}($\vb*{b}_{\rm dev}$,$\vb*{b}$,length($\vb*{b}_{\rm Constraints}$),\texttt{\textbf{cudaMemcpyHostToDevice}})} \; \label{algo:Full_application:copyin_b_aprod1}
		\tcp{Iterative estimate of the solution array $\vb*{x}$}
		\textit{aprod} 2 call (in each MPI process)\; \label{algo:Full_application:aprod2}
		\textcolor{gray}{\texttt{\textbf{cuda\_error = cudaMemcpy}}($\vb*{x}$,$\vb*{x}_{\rm dev}$,length($\vb*{x}$),\texttt{\textbf{cudaMemcpyDeviceToHost}})} \; \label{algo:Full_application:copyout_x_aprod2}
		\tcp{Reduction of $\vb*{x}$ among the MPI processes}
		\texttt{MPI\_Allreduce($\vb*{x}$)}\; \label{algo:Full_application:MPI_reduce_x_aprod2}
		\textcolor{gray}{\texttt{\textbf{cuda\_error = cudaMemcpy}}($\vb*{x}_{\rm dev}$,$\vb*{x}$,length($\vb*{x}$),\texttt{\textbf{cudaMemcpyHostToDevice}})} \; \label{algo:Full_application:copyin_x_aprod2}
		Variances and covariances computation\; \label{algo:Full_application:var_covar}
	} \label{algo:Full_application:LSQR_end}
    Re-multiplication of the solution and of its variance by $\vb*{p}$\; \label{algo:Full_application:precond_3}
    Print the solution to files\;
	
	\caption{Structure of the entire Gaia AVU--GSR application in CUDA \label{algo:Full_application}}
\end{algorithm*}

The coefficient matrix of the system $\mathbf{A}$ is large and has a high sparsity degree. Specifically, for the expected final dataset of Gaia, the matrix might contain $\sim$$10^{11} \times 10^8$ elements (see Sect.~\ref{sec:Intro}). The rows of $\mathbf{A}$, i.e., the equations of the system, represent the observations of the Milky Way stars, where each star is observed $N_{\rm Obsperstar} \sim 10^3$ times, besides the constraints equations. The number of the columns of $\mathbf{A}$ is instead the number of unknowns to solve. 

For each row, the coefficients are divided in their astrometric, attitude, instrumental, and global sections. The astrometric part of $\mathbf{A}$ contains $N_{\rm Astro} \times N_{\rm Stars}$ coefficients per row. $N_{\rm Stars}$ is the number of stars considered in the system, in the range of $[10^6,10^8]$, and $0 \leq N_{\rm Astro} \leq 5$ is the number of astrometric coefficients per star and the number of non-zero astrometric coefficients per row. The total number of non-zero astrometric parameters is of $N_{\rm Obsperstar} \times N_{\rm Astro} \times N_{\rm Stars} \in [10^9,10^{12}]$ and they represent the $\sim$90\% of the coefficient matrix $\mathbf{A}$, where they follow a block-diagonal structure of $N_{\rm Stars}$ blocks. The $N_{\rm Obsperstar}$ rows of each block are the astrometric parameters observed for the same star and the number of columns of each block is equal to $N_{\rm Astro}$. In our current modelization, the attitude part has $N_{\rm Att} = 12$ nonzero coefficients per row, organized in $N_{\rm Axes} = 3$ blocks of $N_{\rm ParAxis} = 4$ elements separated by $N_{\rm DFA}$ zeros, where $N_{\rm Axes} = 3$ is the number of axes of the satellite attitude, $N_{\rm ParAxis} = 4$ is the number of nonzero coefficients per axis, and $N_{\rm DFA}$ is the number of degrees of freedom of each axis. In the instrumental part, we have $0 \leq N_{\rm Instr} \leq 6$ nonzero coefficients per row, distributed without a particular scheme. So far, we have considered in the global part only $N_{\rm Glob} = 1$ coefficient, the $\gamma$ parameter of the PPN formalism, or we have run without computing a global part.

To operate in human-size timescales, the calculations are performed with a dense coefficient matrix $\mathbf{A_{\rm d}}$ that only contains the nonzero coefficients of $\mathbf{A}$ for each section. Therefore, the number of coefficients per row passes from $\sim$$10^8$ to a maximum of $N_{\rm par} = 24$, in our current modelization, and the total number of elements of $\mathbf{A_{\rm d}}$ is of $\sim$$10^{11} \times 10^1$. The indexes that the astrometric, the attitude, and the instrumental coefficients of $\mathbf{A_{\rm d}}$ had in the original matrix $\mathbf{A}$ are stored in two one-dimensional integer arrays, $M_{\rm i}$ (for the astrometric and attitude parts) and $I_{\rm c}$ (for the instrumental part), to map the correct positions of these parameters in the matrix $\mathbf{A}$. For further details about the structure of the coefficient matrix, see Sections 3 and 4 of~\citet{Cesare_2022}.
 
\begin{figure*}[hbt!]
	\centering
	\includegraphics[width=0.95\textwidth]{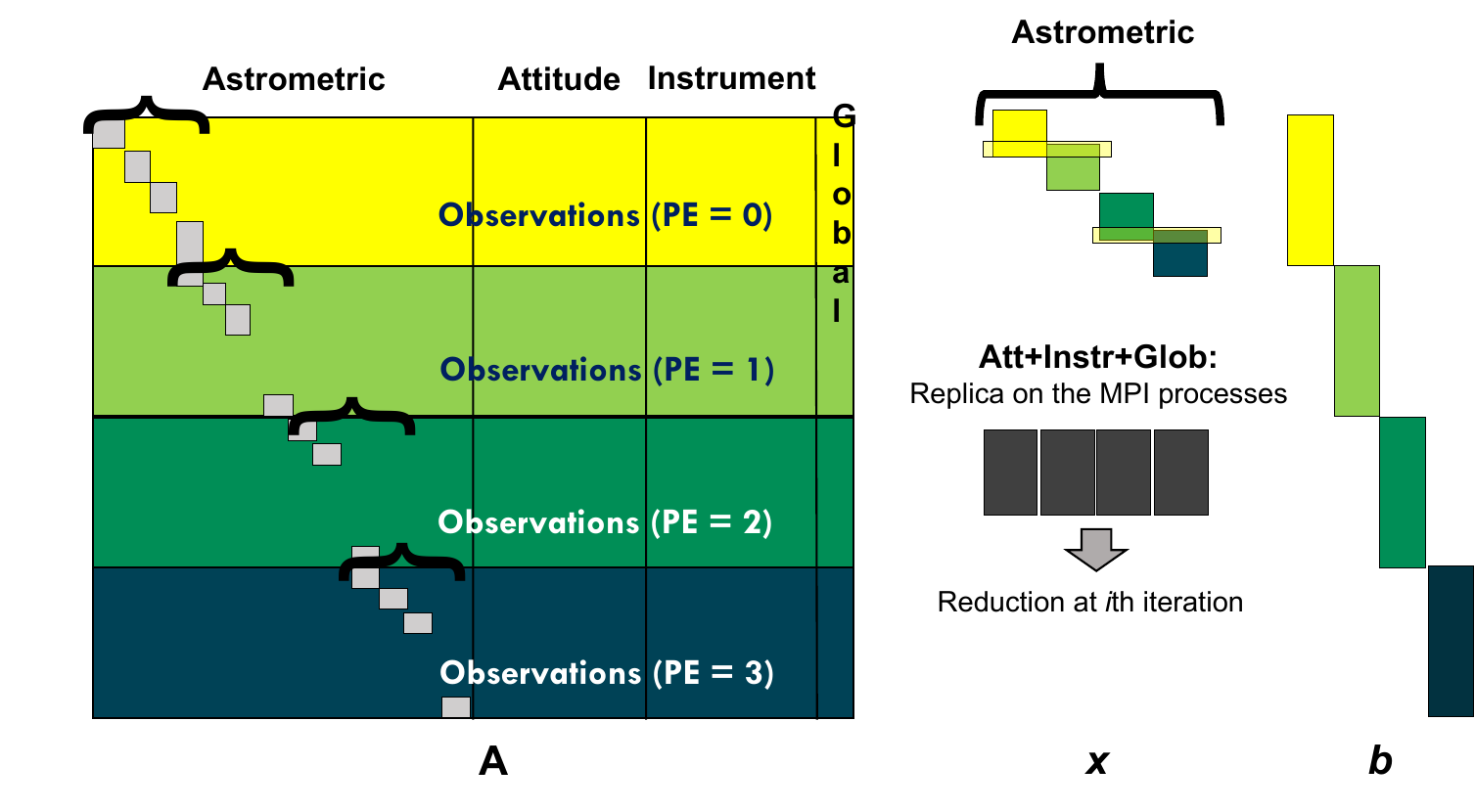}
	\caption{Parallelization scheme of the system of equations (Eq.~\ref{eq:Axb}) on four MPI processes in a single node of a computer cluster. \textit{Left panel}: coefficient matrix $\mathbf{A}$. \textit{Middle panel}: unknowns array $\vb*{x}$. \textit{Right panel}: known terms array $\vb*{b}$. Different colors (yellow, light green, dark green, and blue) refer to different MPI processes or processing elements (PE). The block-diagonal part in the left side of the coefficient matrix illustrates its nonzero astrometric section. In the middle panel, the four square blocks diagonally placed, and labeled as ``Astrometric'' represent the astrometric part of the solution array, distributed among the MPI processes. Instead, the four dark gray aligned blocks, labeled as ``Att+Instr+Glob'', represent the attitude, instrumental, and global portions of the solution array, replicated on each MPI process, as written above. At the end of each iteration $i$, a reduction of the replicated portions of $\vb*{x}$ is performed.}
	\label{fig:Par_Data_schemes}
\end{figure*}

The system of equations is parallelized over the MPI processes such that different subsets of the total number of observations, $n$, are assigned to each MPI process. The one-dimensional integer array  $N[nproc]$ stores the number of observations assigned to each MPI process, where $nproc$ is the number of MPI processes defined at runtime. Figure~\ref{fig:Par_Data_schemes} represents the system of equations parallelized on four MPI processes in one node of a computer cluster, where different colors refer to diverse MPI processes.

Concerning the rows of $\mathbf{A}$, whereas the observation equations are distributed among the MPI processes throught the $N$ array, the constraints equations, placed at the bottom of the system, are replicated on each MPI process. For this reason, after the execution of the \textit{aprod 1} function, the only part of the known terms array $\vb*{b}$ that has to be reduced among the MPI processes is the one related to the constraints equations. Given that the constraints equations represent a negligible fraction of the total number of equations, their replica was more convenient compared to their distribution among the MPI processes, which would have implied a rearrangement of the code.

Concerning the columns of $\mathbf{A}$, the astrometric part is distributed among the MPI processes whereas the other three parts are replicated on them. Therefore, after the execution of the \textit{aprod 2} function, only the attitude + instrumental + global parts of the solution array $\vb*{x}$ are reduced among the MPI processes. The regular block-diagonal structure of the astrometric parameters made their distribution among the MPI processes more intuitive. Instead, the other three sections do not follow a regular pattern, which would have made their distribution on the MPI processes less trivial. Since, the attitude + instrumental + global parts only represent the 10\% of the total system, their replica does not imply a substantial slowdown of the code.

\section{Previous parallelizations: MPI + OpenMP and MPI + OpenACC}
\label{sec:Gaia_AVU_GSR_code_structure_OpenMP_OpenACC}

\subsection{The OpenMP parallelization}
\label{sec:Gaia_AVU_GSR_code_structure_OpenMP}

In the in-production code, the observations assigned to each MPI process are further parallelized over the OpenMP threads. The left panels of Algorithms~\ref{algo:aprod_1} and~\ref{algo:aprod_2} highlights in boldface the regions of the code parallelized with OpenMP, namely the \textit{aprod 1} and \textit{2} functions. In the \textit{aprod 1}, we parallelized the for loop that iterates on the number of observations in each MPI process, $N[pid]$, with the \texttt{\#pragma omp for} directive, where $pid$ identifies the rank of the MPI process. Instead, in the \textit{aprod 2} the most external for loop iterates from $N_{\rm t}[tid][0]$ to $N_{\rm t}[tid][1]$, where $tid$ is the ID number of the OpenMP thread, that goes from 0 to $nth$, the total number of threads, and $N_{\rm t}$ is a one-dimensional integer array that contains the observations computed by each thread $tid$. Specifically, $N_{\rm t}[tid][0]$ and $N_{\rm t}[tid][1]$ are the first and the last observation computed by the thread $tid$.

\subsection{The OpenACC parallelization}
\label{sec:Gaia_AVU_GSR_code_structure_OpenACC}

In our preliminary porting to a GPU environment, the OpenMP parallelization model is replaced by OpenACC~\citep{Cesare_OpenACC_ADASS_XXXI_in_press,Cesare_INAF_Technical_Report_OpenACC_163_2022,Cesare_2022}. The middle panels of Algorithms~\ref{algo:aprod_1} and~\ref{algo:aprod_2} highlights in boldface the correspondent parts of the left panels, parallelized with OpenACC instead of OpenMP. For reasons of optimization, we divided the \textit{aprod 1} function in four parallel regions, one for each section of the system, and we organized the \textit{aprod 2} in a single parallel region. Each parallel region is enclosed within a \texttt{\#pragma acc parallel} directive, which starts a parallel execution on the current device.
In the \textit{aprod 1}, the variable $sum$ is defined within the \texttt{private} clause (lines~\ref{algo:acc:astro_par_sect_beg},~\ref{algo:acc:att_par_sect_beg},~\ref{algo:acc:instr_par_sect_beg}, and~\ref{algo:acc:glob_par_sect_beg} of Algorithm~\ref{algo:aprod_1}), which ensures each GPU thread to have a local copy of it. In both \textit{aprod 1} and \textit{aprod 2}, we parallelized the most external for loop in each parallel region with the \texttt{\#pragma acc loop} directive (lines~\ref{algo:acc:astro_par_loop},~\ref{algo:acc:att_par_loop},~\ref{algo:acc:instr_par_loop}, and~\ref{algo:acc:glob_par_loop}, in Algorithm~\ref{algo:aprod_1}, and line~\ref{algo:acc:aprod_2_ext_loop}, in Algorithm~\ref{algo:aprod_2}). These for loops iterate on the observations related to each MPI process from 0 to $N[pid]$, also in the \textit{aprod 2} function, where the array $N_{\rm t}[$ is no more needed since we do not use any longer the OpenMP threads. In the \textit{aprod 2}, the \texttt{\#pragma acc atomic} directive (lines~\ref{algo:acc:atomic_Astro},~\ref{algo:acc:atomic_Att},~\ref{algo:acc:atomic_Instr}, and~\ref{algo:acc:atomic_Glob}, of Algorithm~\ref{algo:aprod_2}) prevents the GPU threads to simultaneously overwrite the same elements of the array $\vb*{x}$, i.e., it avoids a data race condition. 

\begin{figure*}[hbt!]
	\centering
	\includegraphics[width=0.80\textwidth]{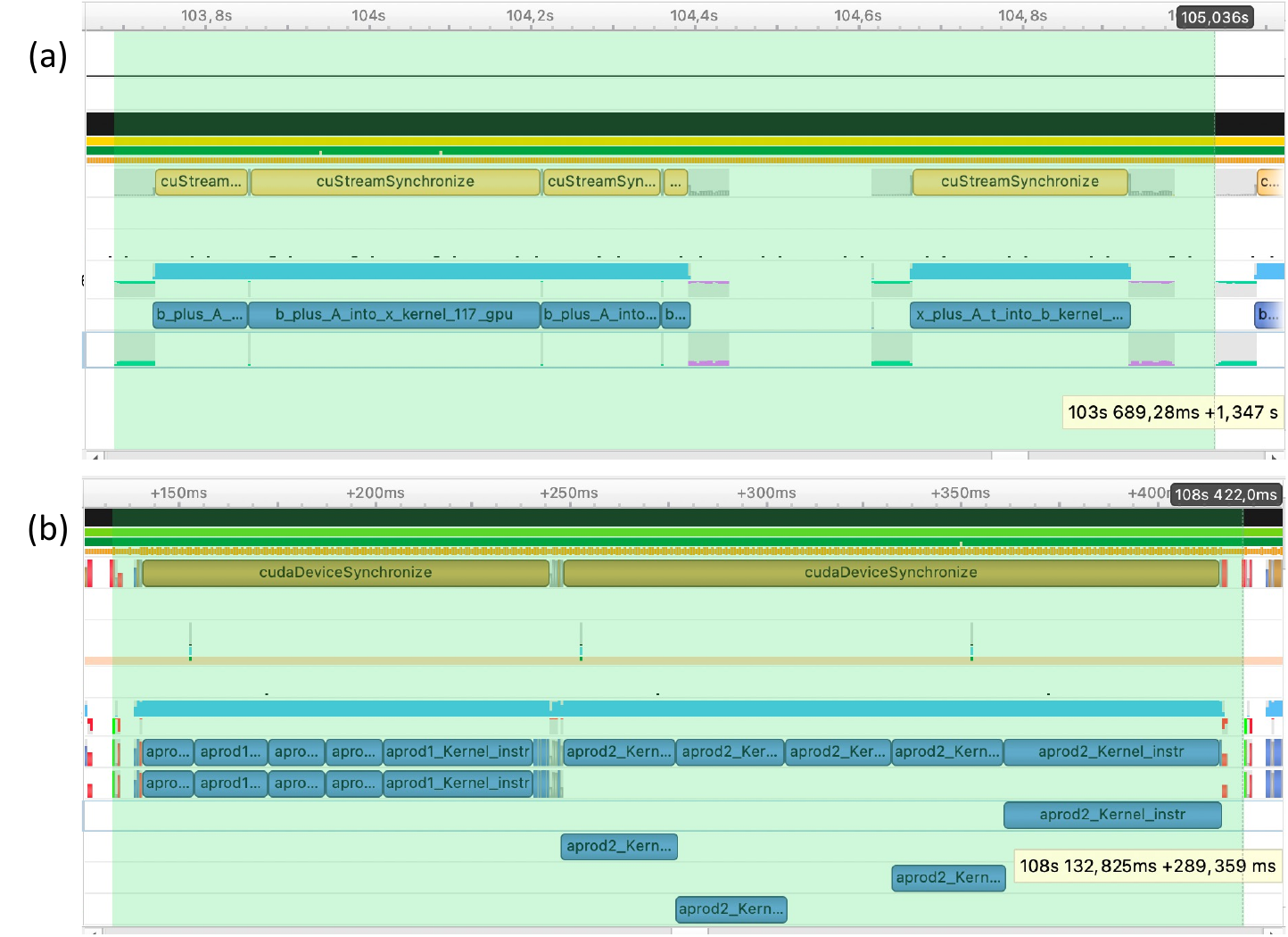}
	\caption{Result of the NVIDIA Nsight Systems profiler for a run of the OpenACC code (\textit{Figure~\ref{fig:Profiler}a}) and of the CUDA code (\textit{Figure~\ref{fig:Profiler}b}) parallelized on 4 MPI processes in one node of M100, for a system that occupies 50 GB of memory. The two outputs show a zoom-in of a single iteration of the LSQR algorithm, better highlighted with a large transparent light green box, and refer to one of the 4 MPI processes. Within the light green box of each panel, the blue regions represent the sections of the code parallelized on the GPU, the green and the purple/red regions illustrate the H2D and D2H data movements (very small in panel~\ref{fig:Profiler}b), and the blank gaps between different regions are related to the sections of code still running on the CPU. The time scale above each panel shows the absolute time from the beginning of the programs execution and the small yellow rectangles at the bottom-right corner of each transparent light green box shows the iteration time (1.347s and 0.289359 s for the OpenACC and the CUDA codes, respectively).}
	\label{fig:Profiler}
\end{figure*}

We tested the performance of the MPI + OpenACC code on M100, with 4 NVIDIA Volta V100 GPUs per node having 16 GB of memory each. Figure~\ref{fig:Profiler}a shows the output of the NVIDIA Nsight System profiler\footnote{\url{https://developer.nvidia.com/nsight-systems}} correspondent to one iteration of the LSQR algorithm, highlighted with a large transparent light green box, for a system occupying 50 GB of memory. The system was run on four MPI processes in one node of M100 and the shown profiler output refers to one of the four processes. The time scale at the top of the panel shows the absolute time from the beginning of the program execution and the small yellow rectangle at the bottom-right corner of the light green box shows the iteration time, equal to 1.347 s. Within the light green box, the profiler shows the code regions parallelized on the GPU (blue), dedicated to data transfers (green and purple, for the H2D and D2H directions), and to calculation on the CPU (blank spaces between different regions). The blue regions labeled as ``\texttt{b\_plus...}'' and as ``\texttt{x\_plus...}'' show the \textit{aprod 1} and \textit{2} functions, respecitvely. The time fractions of one iteration due to GPU computation is $\sim$70\%, whereas the ones due to data transfers and CPU computation are of $\sim$15\%. This shows that the code is \textit{compute bound}, namely data copies are subdominant compared to computation. This is essential for a GPU-ported application that, if data movements are not properly managed, can result in an even worse performance than the correspondent CPU version.

With this parallelization, the OpenACC code accelerates of $\sim$1.5x over the OpenMP version. Specifically, the speedup is due to the porting of the \textit{aprod 2} region, that accelerates of $\sim$3.6x, whereas the \textit{aprod 1} region looses in performance of $\sim$0.8x~\citep{Cesare_2022}. Further optimizations are possible, as better detailed in the following sections.

\section{The CUDA porting of the Gaia AVU--GSR code}
\label{sec:Gaia_AVU_GSR_code_CUDA_porting}

To further improve the performance of the Gaia AVU--GSR solver, we decided to change the parallelization approach. Instead of optimizing the high-level OpenACC parallelization, which might be possible, for example by manually defining the grid of the GPU threads on which each parallel region is run rather than leaving this task to the compiler, we decided to adopt a low-level model, that is the NVDIA native language CUDA. This choice was driven by the fact that, in future further optimizations, this approach will better lead us to manually tune some parameters directly related to the device. In the CUDA code, we manually allocated the GPU threads where to run each parallel region in a grid of threads blocks, each one customized to match the GPU architecture and the topology of the problem to solve.

Whereas the high-level OpenACC porting implied a minimal re-design of the application, ideal for beginner users~\citep{Cesare_2020a,Aldinucci_2021}, at the expense of a possible performance loss, the low-level parallelization with CUDA required a substantial re-engineering of the code structure. This can be seen from the left, middle, and right columns of Algorithms~\ref{algo:aprod_1} and~\ref{algo:aprod_2}, that represent the parallelization of the \textit{aprod 1} and \textit{2} functions with OpenMP, OpenACC, and CUDA, respectively. Comparing the left and the middle columns, we can see that the structure of the OpenMP and of the OpenACC \textit{aprod 1} and \textit{2} functions, both parallelized through high-level directives, are very similar, whereas the right columns of the same algorithms show that the structure of the CUDA \textit{aprod} functions is different.

In the below sections, we detail the parallelization of the CUDA code on multiple GPUs (Section~\ref{sec:Gaia_AVU_GSR_code_CUDA_porting_Multi_GPU}), the definition of the CUDA kernels for the \textit{aprod 1} and \textit{2} functions (Section~\ref{sec:Gaia_AVU_GSR_code_CUDA_porting_CUDA_kernels_Definition}), the GPU porting of regions that in the OpenACC code were still running on the CPU (Section~\ref{sec:Gaia_AVU_GSR_code_CUDA_porting_CPU}), the management of the data-transfers between the host and the device (Section~\ref{sec:Gaia_AVU_GSR_code_CUDA_porting_Data_copies}), and the compilation of the application (Section~\ref{sec:Gaia_AVU_GSR_code_CUDA_porting_Compilation}).

\subsection{Multi-GPU parallelization}
\label{sec:Gaia_AVU_GSR_code_CUDA_porting_Multi_GPU}

As the OpenACC code~\citep{Cesare_INAF_Technical_Report_OpenACC_163_2022,Cesare_OpenACC_ADASS_XXXI_in_press,Cesare_2022}, the CUDA code runs on multiple GPUs, according to the number of MPI processes set at runtime. Specifically, the MPI processes are scheduled on the GPUs of the node in a round-robin fashion. This operation is performed with the commands at lines~\ref{algo:Full_application:CUDA_get_number_GPUs_in_node} and~\ref{algo:Full_application:CUDA_set_number_GPU_in_node} of Algorithm~\ref{algo:Full_application} highlighted in gray. The optimal configuration to run the code is to set the number of MPI processes per node to the number of the GPUs of the node (4 on M100), since it allows to obtain the best performance by exploiting all the GPUs of the node and simultaneously employing the minimal number of MPI resources, as also shown in Section 7.1 of~\citealt{Cesare_2022}.

\subsection{CUDA kernels definition in \textit{aprod 1} and \textit{aprod 2} functions}
\label{sec:Gaia_AVU_GSR_code_CUDA_porting_CUDA_kernels_Definition}

The right columns of Algorithms~\ref{algo:aprod_1} and~\ref{algo:aprod_2} show, in boldface, the definition of the CUDA kernels for the astrometric, the attitude, the instrumental, and the global sections of the \textit{aprod 1} and \textit{2} functions (lines c.\ref{algo:aprod_1:c:CUDA_kernels_definition_start}--\ref{algo:aprod_1:c:CUDA_kernels_definition_end}, in Algorithm~\ref{algo:aprod_1}, and lines c.\ref{algo:aprod_2:c:CUDA_kernels_definition_start}--\ref{algo:aprod_2:c:CUDA_kernels_definition_end} , in Algorithm~\ref{algo:aprod_2}) and their call in the main scope of the program (lines c.\ref{algo_aprod_1:c:kernels_call_in_main_scope_start}--\ref{algo_aprod_1:c:kernels_call_in_main_scope_end}, in Algorithm~\ref{algo:aprod_1}, and lines c.\ref{algo_aprod_2:c:kernels_call_in_main_scope_start}--\ref{algo_aprod_2:c:kernels_call_in_main_scope_end}, in Algorithm~\ref{algo:aprod_2}). The index $i =$ \texttt{blockIdx.x*blockDim.x + threadIdx.x} defined within the kernels, is the global index of the GPU thread within the grid of blocks of threads, where \texttt{blockIdx.x} is the index of the block inside the grid, \texttt{blockDim.x} is the size of the block in threads unit, and \texttt{threadIdx.x} is the thread index local to each block. The arrays involved in the calculations in the GPU kernels, such as the dense system matrix $\mathbf{A}_{\rm d}$, the solution array $\vb*{x}$, and the known terms array $\vb*{b}$, have to be first allocated and then copied on the GPU. In Algorithm~\ref{algo:Full_application}, only the device allocation of $\mathbf{A}_{\rm d}$ is highlighted gray, as an example (line~\ref{algo:Full_application:cudaMalloc_Example}), whereas all the H2D and D2H copies are reported in gray. The arrays allocated on the device are identified with the ``dev'' subscript, as we can see in the kernels in the left columns of Algorithms~\ref{algo:aprod_1} and~\ref{algo:aprod_2}.

Comparing the right and the middle columns of Algorithms~\ref{algo:aprod_1} and~\ref{algo:aprod_2}, we can see that the content of the CUDA kernels that compute the different sections of the system, except for the global part of the \texttt{aprod 2} function, are equivalent to the correspondent parts in the OpenACC code, which are directly defined in the main scope of the program within the for loops iterating on the number of observations assigned to each MPI process $pid$, parallelized with the \texttt{\#pragma acc loop} directive.

To parallelize the for loops iterating on the observations assigned to each MPI process, from observation 0 to observation $N[pid]$, in each section of the system, the index of the GPU thread was directly mapped to the index of the observation. In this way, each thread can independently perform the product $\vb*{b} = \mathbf{A}_{\rm d} \times \vb*{x}$, in the \textit{aprod 1} kernels, and $\vb*{x} = \mathbf{A}^{\rm T}_{\rm d} \times \vb*{b}$, in the \textit{aprod 2} kernels. In these CUDA kernels, the for loop syntax disappears and it is replaced by the if-condition $i < N[pid]$ (lines~c.\ref{algo:aprod_1:c:if_astro}, c.\ref{algo:aprod_1:c:if_att_0}, c.\ref{algo:aprod_1:c:if_instr}, and c.\ref{algo:aprod_1:c:if_glob} of Algorithm~\ref{algo:aprod_1}, and lines~c.\ref{algo:aprod_2:c:if_astro}, c.\ref{algo:aprod_2:c:if_att_0}, and c.\ref{algo:aprod_2:c:if_instr} of Algorithm~\ref{algo:aprod_2}), to avoid the thread index $i$ to point to a memory address beyond the size of the product arrays.  

It is essential to define, in each kernel, the hierarchy of the GPU threads to best match the GPU architecture and the topology of the problem to solve, which implies an efficient exploitation of the hardware of the device and results in an optimal performance. In our case, the topology of the problem is one-dimensional, since the product arrays are 1D. The grid of threads can be defined in a Cartesian coordinate space set by the $x$, $y$, and $z$ axes. The three directions are not equivalent to each other: specifically, along the $x$ direction it is possible to allocate more threads than along the $y$ and the $z$ directions. In particular, on a V100 GPU we can allocate $\sim$$2 \times 10^9$ threads along the $x$ direction and $\sim$$6.5 \times 10^4$ threads along the $y$ and $z$ directions. We thus defined the grids of threads along the $x$ direction, as identified by the \texttt{.x} specification (lines c.\ref{algo:aprod_1:c:grid_x_astro}, c.\ref{algo:aprod_1:c:grid_x_att_0}, c.\ref{algo:aprod_1:c:grid_x_instr}, and c.\ref{algo:aprod_1:c:grid_x_glob} of Algorithm~\ref{algo:aprod_1}, and lines c.\ref{algo:aprod_2:c:grid_x_astro}, c.\ref{algo:aprod_2:c:grid_x_att_0}, and c.\ref{algo:aprod_2:c:grid_x_instr} of Algorithm~\ref{algo:aprod_2}).

As in the OpenACC code, in the \textit{aprod 2} kernels the operations $\vb*{x}$ += $\mathbf{A}^{\rm T}_{\rm d} \times \vb*{b}$ are performed atomically. In CUDA, the atomic operation is performed with the \texttt{atomicAdd} function, that takes as first argument the memory address of the element of the $x$ array where the result is cumulated and as second argument the quantity that has to be cumulated (lines c.\ref{algo:aprod_2:c:atomic_astro}, c.\ref{algo:aprod_2:c:atomic_att_0}, and c.\ref{algo:aprod_2:c:atomic_instr} of Algorithm~\ref{algo:aprod_2}).

Performing different tests, we verified that defining more kernels allows to save the $\sim$10-30\% of the computation time compared to perform more operations in the same kernel. For this reason, we parallelized the four sections of the system both in the \textit{aprod 1} and in the \textit{aprod 2} on more kernels, differently from the \textit{aprod 2} in the OpenACC code, which was defined within a single parallel region. Moreover, we also split the calculation of the attitude section, both in the \textit{aprod 1} and \textit{2} regions, in three kernels, one for each attitude axis. In Algorithms~\ref{algo:aprod_1} and~\ref{algo:aprod_2}, we only report the attitude kernel for axis 0 since the kernels for the other two axes are equivalent.

The parallelization of the global part of the \textit{aprod 2} was less trivial than the other three parts. Looking at the OpenACC column of Algorithm~\ref{algo:aprod_2}, at line~a.\ref{algo:aprod_2:a:global_operation}, we can see that the index of the element of the $\vb*{x}$ array where the result of the atomic operation is cumulated does not depend on the index of the observation $i$, differently from the other three sections (lines~a.\ref{algo:aprod_2:a:astro_operation}, a.\ref{algo:aprod_2:a:att_operation}, a.\ref{algo:aprod_2:a:instr_operation}, c.\ref{algo:aprod_2:c:atomic_astro}, c.\ref{algo:aprod_2:c:atomic_att_0}, and c.\ref{algo:aprod_2:c:atomic_instr} of Algorithm~\ref{algo:aprod_2}). This might cause a bottleneck in this point of the code since there is no parallelism over the GPU threads, whereas, in the astrometric, attitude, and instrumental parts of the \textit{aprod 2}, the access to the element of the $\vb*{x}$ array where the result is cumulated occurs in parallel for each thread $i$ matched to the observation $i$.

So far, for scientific purposes of the current production, we did not derive the $\gamma$ PPN parameter and, thus, this section does not represent a bottleneck. However, the $\gamma$ parameter will be calculated in upcoming runs to test General Relativity, and we cannot exclude that future astrometric models will have more global parameters, which makes necessary to properly parallelize this region of code. To verify how leaving the atomic operation in the \textit{aprod 2} global part would affect the performance of the code, we ran a 6 GB and a 50 GB system with 1 and 5 global parameters. Even with one global parameter, the computation of the \textit{aprod 2} global section dominates over the other sections. We, thus, decided to rearrange the parallelization of this part. We removed the atomic operation and we implemented in CUDA the sum at line~a.\ref{algo:aprod_2:a:global_operation} of Algorithm~\ref{algo:aprod_2} with a parallel reduction, which also exploits the GPU shared memory. This reduce sum is organized in two kernels, the former parallelized on a certain number of blocks, each of which computes a partial sum saved in the array $x_{\rm dev,sum}$ (line c.\ref{algo:aprod_2:c:glob_1}, Algorithm~\ref{algo:aprod_2}), and the latter parallelized on a single block of threads, which combines all the partial results in $x_{\rm dev}$ (line c.\ref{algo:aprod_2:c:glob_2}, Algorithm~\ref{algo:aprod_2}). Adopting this new implementation, we obtain a $\sim$20x speedup for the global part of the \textit{aprod 2} function and a speedup of $[1.5,3]$x for the entire aprod 2 region. A deeper description of this implementation for the \textit{aprod 2} global part will be object of a future work.

The grid of threads on which the kernels are parallelized, except for the \textit{aprod 2} global kernels, are set through the \texttt{gridDim} and \texttt{blockDim} vectors, defined with the \texttt{dim3} integer vector type. The \texttt{gridDim} vector contains three elements, i.e., the number of blocks of threads in the grid along the $x$, $y$, and $z$ directions. Since we defined a 1D grid, we have 1 block along the $y$ and $z$ directions. Instead, the \texttt{blockDim} vector contains the number of threads in each block again along the $x$, $y$, and $z$ directions. As in \texttt{gridDim}, the second and the third element of \texttt{blockDim} are set to 1.

On a V100 GPU, the maximum number of threads per block is 1024. We set the number of threads per block, ${\rm} TW$, to 1024, since it allows to obtain the best performance.
Since the regions that we parallelized with the CUDA kernels correspond to the for loops that iterate from observation 0 to observation $N[pid]$ and that the index $i$ of the observation is directly mapped to the index $i$ of the GPU thread, the number of blocks should be $N[pid]/{\rm TW}$ to properly fit the problem. If $N[pid]$ were an exact multiple of ${\rm TW}$, this would result in a grid of $N[pid]$ threads. Yet, this is not necessary the case, and to avoid defining a grid with less threads than required, the number of blocks is set to $N_{\rm bl} = (N[pid] - 1)/{\rm TW} + 1$. Since the total number of threads $N_{\rm th} = N_{\rm bl} \times {\rm TW} = N[pid] + {\rm TW} -1  > N[pid]$, the control condition ${\rm if} i < n$ defined in the kernels (lines~c.\ref{algo:aprod_1:c:if_astro}, c.\ref{algo:aprod_1:c:if_att_0}, c.\ref{algo:aprod_1:c:if_instr}, and c.\ref{algo:aprod_1:c:if_glob} of Algorithm~\ref{algo:aprod_1} and lines c.\ref{algo:aprod_2:c:if_astro}, c.\ref{algo:aprod_2:c:if_att_0}, and c.\ref{algo:aprod_2:c:if_instr} of Algorithm~\ref{algo:aprod_2}) is necessary to avoid memory overflows. The scalar $n$ coincides with $N[pid]$, as specified by the parameter $N[pid]$ passed to the kernel at lines~c.\ref{algo_aprod_1:c:kernels_call_in_main_scope_start}-c.\ref{algo_aprod_1:c:kernels_call_in_main_scope_end} of Algorithm~\ref{algo:aprod_1} and at lines~c.\ref{algo_aprod_2:c:kernels_call_in_main_scope_start}-c.\ref{algo_aprod_2:c:kernels_call_in_main_scope_before_global_part_end} of Algorithm~\ref{algo:aprod_2}. For the \textit{aprod 2} global kernels, we defined a different grid (lines c.\ref{algo_aprod_2:c:grid_def_aprod_2_glob_1} and c.\ref{algo_aprod_2:c:grid_def_aprod_2_glob_2} of Algorithm~\ref{algo:aprod_2}), where the number of threads per block is always set to 1024 and the number of blocks employed for the reduction operation is $24 \times 16$, empirically obtained such that it provided the best performance.

In the kernels call within the main scope of the program, the parameters passed in the angle brackets are the number of blocks and of threads per block. In the \textit{aprod 2}, two additional arguments are passed within the angle brackets of the astrometric, the attitude, and the instrumental kernels. The third parameter sets the amount of the GPU shared memory to be used by the kernel, in this case 0 bytes, and the fourth argument is the identifier of the execution queue, called stream, on the GPU. This allows to execute these kernels asynchronously, which is possible without imparing the code correctness due to the atomic operations. The overlapping execution regions between the different \textit{aprod 2} kernels represent a very minor fraction of the kernels execution, since the number of threads that can concurrently run on the GPU is limited to the number of GPU cores, much smaller than the number of threads in the grid. Yet, the asyncronous computation is useful to reduce the latencies between the successive kernels calls, essential when a large number of iterations is required to reach the convergence of the LSQR algorithm.

After the call of all the CUDA kernels of the \textit{aprod 1} and \textit{aprod 2} regions, a \texttt{cudaDeviceSynchronize()} barrier is set to wait all the kernels to end their computation (line c.\ref{algo:aprod_1:c:cudaDeviceSynchronize} of Algorithm~\ref{algo:aprod_1} and line c.\ref{algo:aprod_2:c:cudaDeviceSynchronize} of Algorithm~\ref{algo:aprod_2}). This is necessary since, as soon as a CUDA kernel is called, the control returns immediately to the host and the CPU operations defined immediately after the kernels calls, i.e., the MPI reduction operations that combine the partial results obtained from the \textit{aprod 1} and \textit{aprod 2} (line c.\ref{algo:aprod_1:c:Reduce} of Algorithm~\ref{algo:aprod_1} and line c.\ref{algo:aprod_2:c:Reduce} of Algorithm~\ref{algo:aprod_2}),  would run concurrently to the calculations performed by the kernels.

Figure~\ref{fig:Profiler}b shows the output of the NVIDIA Nsight System profiler for a 50 GB run of the CUDA code correspondent to the one of Figure~\ref{fig:Profiler}a for the OpenACC code, i.e., parallelized on 4 MPI processes in one node of M100. As for Figure~\ref{fig:Profiler}a, the top time scale refers to the absolute time from the start of the program execution and the value in the yellow rectangle shows the time for the illustrated iteration (0.289359 s). Comparing the outputs of Figures~\ref{fig:Profiler}a and~\ref{fig:Profiler}b, the \textit{aprod 1} and \textit{2} regions computed with the CUDA code present a $6.4$x and $1.6$x speedup over the same sections computed with the OpenACC application, respectively, correspondent to a speedup of $5.1$x and $5.8$x over the OpenMP code for an analogous run~\citep{Cesare_INAF_Technical_Report_OpenACC_163_2022,Cesare_2022}. Considering one entire iteration, the CUDA code accelerates of $\sim$5x over the OpenACC code and of $\sim$7x over the OpenMP code. However, the speedup increases with the memory occupied by the system and with the resources employed for the parallelization, as better illustrated in Section~\ref{sec:Gaia_AVU_GSR_code_performance_tests}.

\subsection{GPU porting of the CPU regions of the code}
\label{sec:Gaia_AVU_GSR_code_CUDA_porting_CPU}

Besides optimizing the parallelization of the \textit{aprod 1} and \textit{aprod 2} regions, we also ported to the GPU other sections of code that in the OpenACC version were running on the CPU. Looking at Figure~\ref{fig:Profiler}a, we can clearly see that many code regions were still running on the CPU (blank gaps). 

First of all, we ported to the GPU the computation of the constraints equations both for the \textit{aprod 1} and the \textit{aprod 2} functions. With this porting, the computation time of these regions remains basically unaffected, since their calculation was already very fast ($\sim$$10^{-4}$~s) on the CPU. However, porting these regions is essential to reduce the H2D and D2H data copies, since it avoids to entirely copy the $\vb*{b}$ and the $\vb*{x}$ arrays back to the host to perform the same calculations on the CPU.

Second of all, we ported to the GPU the calculation of the quantity to be compared with the tolerance of $10^{-16}$ that determines the convergence of the LSQR algorithm. This quantity depends on the norms, $\beta$ and $\alpha$, of the $\vb*{b}$ and the $\vb*{x}$ arrays. In the AVU--GSR code, the norm of these two arrays is calculated, in each MPI process, by square summing the array elements divided by the array maximum local to the MPI process and by multiplying this normalized squared sum by this maximum:
\begin{equation}
\label{eq:Norm_loc_ssq}
\vert \vb*{a} \vert_{loc} = max\text{\_}loc(a) \times \sqrt{\sum_{i=0}^{N}\frac{a^2_i}{max\text{\_}loc(a)^2}},
\end{equation} 
where $\vb*{a}$ is either $\vb*{b}$ or $\vb*{x}$. Then, the global norm of the array, $\vert \vb*{a} \vert$, is reduced among the MPI processes and sent back to each MPI process.

This method to compute the norm is adopted not to loose in numerical precision. In the MPI + OpenMP and MPI + OpenACC codes, this norm is computed on the CPU with the \texttt{cblas\_dnrm2} function of the cblas libraries~\citep{Galassi_GSL_2018}\footnote{http://www.gnu.org/software/gsl/}.
In the MPI + CUDA version, we ported to the GPU the calculation of this norm by computing the local maximum and the squared and scaled sum of the array elements with a parallel reduce operation. 
With this porting, the computation of $\beta$ and $\alpha$ accelerates of $\sim$$35$x over the CPU implementation. 

By porting these calculations to the GPU, the time fraction of one iteration due to CPU computation reduces from $\sim$15\% (see Section~\ref{sec:Gaia_AVU_GSR_code_structure_OpenACC}), to $\sim$3\%, and the time fraction due to GPU computation increases from $\sim$70\% (see Section~\ref{sec:Gaia_AVU_GSR_code_structure_OpenACC}) to $\sim$90\%. This can be visually seen from Figure~\ref{fig:Profiler}b, where the blank regions of the profiler are drastically reduced compared to Figure~\ref{fig:Profiler}a.

\subsection{H2D and D2H data transfers}
\label{sec:Gaia_AVU_GSR_code_CUDA_porting_Data_copies}

In the OpenACC code, we copied, at each iteration, both the entire $\vb*{b}$ and $\vb*{x}$ arrays, from the host to the device before the beginning of the \textit{aprod 1} and \textit{aprod 2} functions, and from the device to the host, after the end of the same functions. The $\vb*{b}$ and the $\vb*{x}$ arrays only represent the 5\% of the total memory occupied by the system of equations~\citep{Cesare_2022} and consequently the time fraction of one iteration due to data copies was anyway subdominant compared to the time fraction due to GPU computation (see Section~\ref{sec:Gaia_AVU_GSR_code_structure_OpenACC} and Figure~\ref{fig:Profiler}a). However, some of these copies were unnecessary and they could be further reduced.

In the CUDA code, the first copy of the entire $\vb*{b}$ and $\vb*{x}$ arrays on the device is performed before the beginning of the LSQR cycle (lines~\ref{algo:Full_application:copyin_b_before_LSQR} and~\ref{algo:Full_application:copyin_x_aprod2_IS}  of Algorithm~\ref{algo:Full_application}). Since the \textit{aprod 1} only modifies the $\vb*{b}$ array, we only copy back to the host this array after the execution of the \textit{aprod 1}, necessary operation since the $\vb*{b}$ array has to be reduced among the MPI processes. In fact, only the constraints part of the $\vb*{b}$ array has to be reduced among the MPI processes (see Section~\ref{sec:Gaia_AVU_GSR_code_structure}). For this reason, we only copy back to the host the final portion of the $\vb*{b}$ array correspondent to the constraints part (see ``length($\vb*{b}_{\rm Constraints}$)'' in the \texttt{cudaMemcpy} commands at lines~\ref{algo:Full_application:copyout_b_aprod1} and~\ref{algo:Full_application:copyin_b_aprod1} of Algorithm~\ref{algo:Full_application}), which represents a minor fraction of the entire array. The same portion of the $\vb*{b}$ array is again copied to the device after the reduction operation. 

For the same reason, since the \textit{aprod 2} only modifies the $\vb*{x}$ array, the $\vb*{x}$ array alone is copied back to the host after the execution of the \textit{aprod 2}. After the copy, the $\vb*{x}$ array is reduced among the MPI processed on the host and then is again copied to the device.

Rearranging the data copies in this way, the time fraction of one iteration due to data copies reduces from $\sim$15\%, in the OpenACC code, to $\sim$3\%, in the CUDA code (see Section~\ref{sec:Gaia_AVU_GSR_code_structure_OpenACC} and Figure~\ref{fig:Profiler}). The data copies in the CUDA code are highlighted in bold gray in Algorithm~\ref{algo:Full_application}.

\subsection{Compilation}
\label{sec:Gaia_AVU_GSR_code_CUDA_porting_Compilation}

To compile the MPI + CUDA code, written both in C and C++, we wrote a Makefile and we employed the nvcc CUDA compiler driver for the CUDA release 11.3 and the version 21.5-0 of the nvc++ compiler. We compile with the -arch=sm\_70 option to target the Volta architecture of the GPU, present on M100. 

\begin{algorithm*}
	\tiny
	\begin{multicols}{3}
		\nonl \textbf{\small \textit{aprod} 1 with OpenMP}\;
		\SetNlSty{}{o.}{}
		int main(int argc, char **argv) \;
		\{ \;
		... \;
		\texttt{\textbf{\#pragma omp parallel private($pid$,$sum$) shared($\vb*{N}$,$\vb*{x}$,$\mathbf{A}_\textbf{d}$,$\vb*{b}$)}} \;
		\{\;
		\text{ }\text{ }\text{ }\texttt{\textbf{\#pragma omp for}}\;
		\text{ }\text{ }\text{ }\For{$i\leftarrow 0$ \KwTo $\vb*{N}[pid]$}  { \label{algo:omp:N_pid}
			$sum = 0.0$\;
			\tcp{Astrometric sect.}
			$k = i \times N_{\rm par}$ \;
			\For{$j\leftarrow 0$ \KwTo $N_{\rm Astro}$}{
				$sum = sum + \mathbf{A}_\textbf{d}[k]\vb*{x}[j + {\rm offset}[i]]$\;
				$k$++\;
			}
			\tcp{Attitude sect.}
			$k = i \times N_{\rm par} + N_{\rm Astro}$ \;
			\For{$j_1\leftarrow 0$ \KwTo $N_{\rm Axes}$}{
				$k_2 = j_1 \times N_{\rm DFA} + {\rm offset}[i] $\;
				\For{$j_2 \leftarrow 0$ \KwTo $N_{\rm ParAxis}$}{
					$sum = sum + \mathbf{A}_\textbf{d}[k]\vb*{x}[j_2 + k_2]$\;
					$k$++\;
				}  
			}
			\tcp{Instrumental sect.} 	        
			$k = i \times N_{\rm par} + N_{\rm Astro} + N_{\rm Att}$ \;
			\For{$j\leftarrow 0$ \KwTo $N_{\rm Instr}$}{
				$sum = sum + \mathbf{A}_\textbf{d}[k]\vb*{x}[\mathcal{F}(i,j) + {\rm offset}]$\;
				$k$++\;
			}
			\tcp{Global sect.}
			$k = i \times N_{\rm par} + N_{\rm Astro} + N_{\rm Att} + N_{\rm Instr}$ \;
			\For{$j\leftarrow 0$ \KwTo $N_{\rm Glob}$}{
				$sum = sum + \mathbf{A}_\textbf{d}[k]\vb*{x}[j + {\rm offset}]$\;
				$k$++\;
			}
			$\vb*{b}[i]$ = $\vb*{b}[i]$ + $sum$\; 
		}
		\}\;
		Constraints computation\;
		\texttt{MPI\_Allreduce($\vb*{b}$)} \; \label{algo:aprod_1:o:Reduce}
		... \;
		\} \;
		\nonl \;
		\nonl \;
		\nonl \;
		\nonl \;
		\nonl \;
		\nonl \;
		\nonl \;
		\nonl \;
		\nonl \;
		\nonl \;
		\nonl \;
		\nonl \;
		\nonl \;
		\nonl \;
		\nonl \;
		\nonl \;
		\nonl \;
		\nonl \;
		\nonl \;
		\nonl \;
		\nonl \;
		\nonl \;
		\nonl \;
		\nonl \;
		\nonl \;
		\nonl \;
		\nonl \;
		\nonl \;
		\nonl \;
		\nonl \;
		\nonl \;
		\nonl \;
		\nonl \;
		\nonl \;
		\nonl \;
		\nonl \;
		\nonl \;
		\nonl \;
		\nonl \;
		\nonl \;
		\nonl \;
		\nonl \;
		\nonl \;
		\nonl \;
		\nonl \;
		\nonl \;
		\nonl \;
		\nonl \;
		\nonl \;
		\nonl \;
		\nonl \;
		\nonl \;
		\nonl \;
		\nonl \;
		\nonl \;
		\nonl \;
		\nonl \;
		\nonl \;
		\nonl \;
		\nonl \;
		\nonl \;
		\nonl \;
		\nonl \;
		\nonl \;
		\nonl \;
		\nonl \;
		\nonl \;
		\nonl \;
		\nonl \;
		\nonl \;
		\nonl \;
		\nonl \;
		\nonl \;
		\nonl \;
		\nonl \;
		
		
		\setcounter{AlgoLine}{0}
		\SetNlSty{}{a.}{}
		\nonl \textbf{\small \textit{aprod} 1 with OpenACC}\;
		int main(int argc, char **argv) \;
		\{ \;
		... \;
		\tcp{Astrometric sect.}
		\texttt{\textbf{\#pragma acc parallel private($sum$)}} \; \label{algo:acc:astro_par_sect_beg}
		\{\;
		\text{ }\text{ }\text{ }\texttt{\textbf{\#pragma acc loop}} \; \label{algo:acc:astro_par_loop}
		\text{ }\text{ }\text{ }\For{$i\leftarrow 0$ \KwTo $\vb*{N}[pid]$}  {
			$sum = 0.0$\;
			$k = i \times N_{\rm par}$\;
			\For{$j\leftarrow 0$ \KwTo $N_{\rm Astro}$}{
				$sum = sum + \mathbf{A}_\textbf{d}[k + j]\vb*{x}[j + {\rm offset}[i]]$\;
				\nonl \; 
			}
			$\vb*{b}[i]$ = $\vb*{b}[i]$ + $sum$\; 
		}
		\}\; \label{algo:acc:astro_par_sect_end}
		\tcp{Attitude sect.}
		\texttt{\textbf{\#pragma acc parallel private($sum$)}} \; \label{algo:acc:att_par_sect_beg}
		\{\;
		\text{ }\text{ }\text{ }\texttt{\textbf{\#pragma acc loop}} \; \label{algo:acc:att_par_loop}
		\text{ }\text{ }\text{ }\For{$i\leftarrow 0$ \KwTo $\vb*{N}[pid]$}  {
			$sum = 0.0$\;
			$k = i \times N_{\rm par} + N_{\rm Astro}$\;
			\For{$j_1\leftarrow 0$ \KwTo $N_{\rm Axes}$}{
				$k_1 = j_1 \times N_{\rm ParAxis} $\;
				$k_2 = j_1 \times N_{\rm DFA} + {\rm offset}[i] $\;
				\For{$j_2 \leftarrow 0$ \KwTo $N_{\rm ParAxis}$}{
					$sum = sum + \mathbf{A}_\textbf{d}[k + j_2 + k_1]\vb*{x}[j_2 + k_2]$\;
				}  
			}
			$\vb*{b}[i]$ = $\vb*{b}[i]$ + $sum$\; 
		}
		\}\; \label{algo:acc:att_par_sect_end}
		\tcp{Instrumental sect.}
		\texttt{\textbf{\#pragma acc parallel private($sum$)}} \; \label{algo:acc:instr_par_sect_beg}
		\{\;
		\text{ }\text{ }\text{ }\texttt{\textbf{\#pragma acc loop}} \; \label{algo:acc:instr_par_loop}
		\text{ }\text{ }\text{ }\For{$i\leftarrow 0$ \KwTo $\vb*{N}[pid]$}  {
			$sum = 0.0$\;
			$k = i \times N_{\rm par} + N_{\rm Astro} + N_{\rm Att}$\;
			\For{$j\leftarrow 0$ \KwTo $N_{\rm Instr}$}{
				$sum = sum + \mathbf{A}_\textbf{d}[k + j]\vb*{x}[\mathcal{F}(i,j) + {\rm offset}]$\;
				\nonl \; 
			}
			$\vb*{b}[i]$ = $\vb*{b}[i]$ + $sum$\; 
		}
		\}\; \label{algo:acc:instr_par_sect_end}
		\tcp{Global sect.}
		\texttt{\textbf{\#pragma acc parallel private($sum$)}} \; \label{algo:acc:glob_par_sect_beg}
		\{\;
		\text{ }\text{ }\text{ }\texttt{\textbf{\#pragma acc loop}} \; \label{algo:acc:glob_par_loop}
		\text{ }\text{ }\text{ }\For{$i\leftarrow 0$ \KwTo $\vb*{N}[pid]$}  {
			$sum = 0.0$\;
			$k = i \times N_{\rm par} + N_{\rm Astro} + N_{\rm Att} + N_{\rm Instr}$\;
			\For{$j\leftarrow 0$ \KwTo $N_{\rm Glob}$}{
				$sum = sum + \mathbf{A}_\textbf{d}[k+j]\vb*{x}[j + {\rm offset}]$\;
				\nonl \; 
			}
			$\vb*{b}[i]$ = $\vb*{b}[i]$ + $sum$\;  
		}
		\}\; \label{algo:acc:glob_par_sect_end}
		\nonl \;  
		Constraints computation\;
		\texttt{MPI\_Allreduce($\vb*{b}$)} \; \label{algo:aprod_1:a:Reduce}
		... \;
		\} \;
		
		\nonl \;
		\nonl \;
		\nonl \;
		\nonl \;
		\nonl \;
		\nonl \;
		\nonl \;
		\nonl \;
		\nonl \;
		\nonl \;
		\nonl \;
		\nonl \;
		\nonl \;
		\nonl \;
		\nonl \;
		\nonl \;
		\nonl \;
		\nonl \;
		\nonl \;
		\nonl \;
		\nonl \;
		\nonl \;
		\nonl \;
		\nonl \;
		\nonl \;
		\nonl \;
		\nonl \;
		\nonl \;
		\nonl \;
		\nonl \;
		\nonl \;
		\nonl \;
		\nonl \;
		\nonl \;
		\nonl \;
		\nonl \;
		\nonl \;
		\nonl \;
		\nonl \;
		\nonl \;
		\nonl \;
		\nonl \;
		\nonl \;
		\nonl \;
		\nonl \;
		\nonl \;
		\nonl \;
		\nonl \;
		\nonl \;
		\nonl \;
		\nonl \;
		\nonl \;
		\nonl \;
		
		\setcounter{AlgoLine}{0}
		\SetNlSty{}{c.}{}
		
		\nonl \textbf{\small \textit{aprod} 1 with CUDA}\;
		
		\tcp{Astrometric sect.}
		\textbf{\texttt{\_\_global\_\_ void} \texttt{aprod1\_Kernel\_astro} \; \nonl (double* $\mathbf{A}_\textbf{d,dev}$, double* $\vb*{x}_\textbf{dev}$,\; \nonl double* $\vb*{b}_\textbf{dev}$, long $n$, \; \nonl short $N_{\rm Astro}$, long $N_{\rm par}$, ...)} \; \label{algo:aprod_1:c:CUDA_kernels_definition_start}
		\{ \;
		\text{ }  \textbf{$i$ = \texttt{blockIdx.x*blockDim.x+threadIdx.x}}\; \label{algo:aprod_1:c:grid_x_astro}
		\text{ }  {\bf if} ($i < n$) \;  \label{algo:aprod_1:c:if_astro}
		\text{ }  \{ \;
		\text{ }\text{ }  $sum = 0.0$\;
		\text{ }\text{ }  $k = i \times N_{\rm par}$\;
		\text{ }\text{ }  \For{$j\leftarrow 0$ \KwTo $N_{\rm Astro}$}{
			$sum$ $+$$=$ $\mathbf{A}_\textbf{d,dev}[k + j]\vb*{x}_\textbf{dev}[j + {\rm offset}[i]]$\;
		}
		\text{ }\text{ }  $\vb*{b}_\textbf{dev}[i]$ = $\vb*{b}_\textbf{dev}[i]$ + $sum$\; 
		\text{ }  \} \;
		\} \;
		
		\nonl \;
		
		\tcp{Attitude sect., axis 0}
		\textbf{\texttt{\_\_global\_\_ void} \texttt{aprod1\_Kernel\_attA0} \; \nonl (double* $\mathbf{A}_\textbf{d,dev}$, double* $\vb*{x}_\textbf{dev}$,\; \nonl double* $\vb*{b}_\textbf{dev}$, long $n$, \; \nonl short $N_{\rm DFA}$, short $N_{\rm ParAxis}$, long $N_{\rm par}$, short $N_{\rm Astro}$ ...)} \;
		\{\;
		
		\text{ } $k_1 = 0 \times N_{\rm ParAxis} $\;
		
		\text{ } \textbf{$i$ = \texttt{blockIdx.x*blockDim.x+threadIdx.x}}\; \label{algo:aprod_1:c:grid_x_att_0}
		\text{ } {\bf if} ($i < n$) \; \label{algo:aprod_1:c:if_att_0}
		\text{ } \{ \;
		\text{ }\text{ } $sum = 0.0$\;
		\text{ }\text{ } $k = i \times N_{\rm par} + N_{\rm Astro}$\;
		\text{ }\text{ } $k_2 = 0 \times N_{\rm DFA} + {\rm offset}[i] $\;
		\text{ }\text{ } \For{$j_2 \leftarrow 0$ \KwTo $N_{\rm ParAxis}$} {
			$sum$ $+$$=$ $\mathbf{A}_\textbf{d,dev}[k + j_2 + k_1]\vb*{x}_\textbf{dev}[j_2 + k_2]$\;
		}
		\text{ }\text{ } $\vb*{b}_\textbf{dev}[i]$ = $\vb*{b}_\textbf{dev}[i]$ + $sum$\; 
		\text{ } \} \;
		\}\;
		
		\tcp{Attitude sect., axis 1}
		\textbf{\texttt{\_\_global\_\_ void} \texttt{aprod1\_Kernel\_attA1}(...)} \; 
		\{...\} \;
		
		\tcp{Attitude sect., axis 2}
		\textbf{\texttt{\_\_global\_\_ void} \texttt{aprod1\_Kernel\_attA2}(...)} \; 
		\{...\} \;
		
		\tcp{Instrumental sect.}
		
		\textbf{\texttt{\_\_global\_\_ void} \texttt{aprod1\_Kernel\_instr} \; \nonl (double* $\mathbf{A}_\textbf{d,dev}$, double* $\vb*{x}_\textbf{dev}$,\; \nonl double* $\vb*{b}_\textbf{dev}$, long $n$, \; \nonl short $N_{\rm Astro}$, short $N_{\rm Att}$, short $N_{\rm Instr}$, long $N_{\rm par}$, ...)} \;
		\{\;
		
		\text{ } \textbf{$i$ = \texttt{blockIdx.x*blockDim.x+threadIdx.x}}\; \label{algo:aprod_1:c:grid_x_instr}
		\text{ } {\bf if} ($i < n$)  \; \label{algo:aprod_1:c:if_instr}
		\text{ } \{ \;
		\text{ }\text{ } $sum = 0.0$\;
		\text{ }\text{ } $k = i \times N_{\rm par} + N_{\rm Astro} + N_{\rm Att}$\;
		\text{ }\text{ } \For{$j\leftarrow 0$ \KwTo $N_{\rm Instr}$} {
			$sum = sum + \mathbf{A}_\textbf{d,dev}[k + j]\vb*{x_\textbf{dev}}[\mathcal{F}(i,j) + {\rm offset}]$\;
			\nonl \; 
		}
		\text{ }\text{ }$\vb*{b}_\textbf{dev}[i]$ = $\vb*{b}_\textbf{dev}[i]$ + $sum$\; 
		\text{ } \} \;
		\} \;
		
		\tcp{Global sect.}
		
		\textbf{\texttt{\_\_global\_\_ void} \texttt{aprod1\_Kernel\_glob} \; \nonl (double* $\mathbf{A}_\textbf{d,dev}$, double* $\vb*{x}_\textbf{dev}$,\; \nonl double* $\vb*{b}_\textbf{dev}$, long $n$, \nonl short $N_{\rm Astro}$,\;  short $N_{\rm Att}$, short $N_{\rm Instr}$, short $N_{\rm Glob}$, long $N_{\rm par}$, ...)} \;
		\{\;
		
		\text{ } \textbf{$i$ = \texttt{blockIdx.x*blockDim.x+threadIdx.x}}\; \label{algo:aprod_1:c:grid_x_glob}
		\text{ } {\bf if} ($i < n$) \; \label{algo:aprod_1:c:if_glob}
		\text{ } \{ \;
		\text{ } \text{ } $sum = 0.0$\;
		\text{ } \text{ }  $k = i \times N_{\rm par} + N_{\rm Astro} + N_{\rm Att} + N_{\rm Instr}$\;
		\text{ } \text{ } \For{$j\leftarrow 0$ \KwTo $N_{\rm Glob}$}{
			$sum = sum + \mathbf{A}_\textbf{d,dev}[k+j]\vb*{x}_\textbf{dev}[j + {\rm offset}]$\;
			\nonl \; 
		}
		\text{ } \text{ } $\vb*{b}_\textbf{dev}[i]$ = $\vb*{b}_\textbf{dev}[i]$ + $sum$\;  
		\text{ } \} \;
		\}\; 
		
		\texttt{\textbf{Constraints kernels}}\; \label{algo:aprod_1:c:CUDA_kernels_definition_end}
		
		... \;
		
		int main(int argc, char **argv) \;
		\{ \;
		... \;
		
		TW = 1024 \;
		\texttt{dim3} gridDim (($N[pid]$ - 1)/TW + 1,1,1) \;
		\texttt{dim3} blockDim (TW,1,1) \;
		\texttt{\textbf{aprod1\_Kernel\_astro<<<}}gridDim,blockDim\texttt{\textbf{>>>}} \; \nonl ($\mathbf{A}_\textbf{d,dev}$,$\vb*{x}_\textbf{dev}$,$\vb*{b}_\textbf{dev}$,$N[pid]$,$N_{\rm Astro}$,$N_{\rm par}$, ...) \; \label{algo_aprod_1:c:kernels_call_in_main_scope_start}
		\texttt{\textbf{aprod1\_Kernel\_attA0<<<}}gridDim,blockDim\texttt{\textbf{>>>}} \; \nonl ($\mathbf{A}_\textbf{d,dev}$,$\vb*{x}_\textbf{dev}$,$\vb*{b}_\textbf{dev}$,$N[pid]$,$N_{\rm DFA}$,$N_{\rm ParAxis}$, $N_{\rm par}$,$N_{\rm Astro}$ ...) \;
		\texttt{\textbf{aprod1\_Kernel\_attA1<<<}}gridDim,blockDim\texttt{\textbf{>>>}} \; \nonl ($\mathbf{A}_\textbf{d,dev}$,$\vb*{x}_\textbf{dev}$,$\vb*{b}_\textbf{dev}$,$N[pid]$,$N_{\rm DFA}$,$N_{\rm ParAxis}$, $N_{\rm par}$,$N_{\rm Astro}$ ...) \;
		\texttt{\textbf{aprod1\_Kernel\_attA2<<<}}gridDim,blockDim\texttt{\textbf{>>>}} \; \nonl ($\mathbf{A}_\textbf{d,dev}$,$\vb*{x}_\textbf{dev}$,$\vb*{b}_\textbf{dev}$,$N[pid]$,$N_{\rm DFA}$,$N_{\rm ParAxis}$, $N_{\rm par}$,$N_{\rm Astro}$ ...) \;
		\texttt{\textbf{aprod1\_Kernel\_instr<<<}}gridDim,blockDim\texttt{\textbf{>>>}} \; \nonl ($\mathbf{A}_\textbf{d,dev}$,$\vb*{x}_\textbf{dev}$,$\vb*{b}_\textbf{dev}$,$N[pid]$,$N_{\rm Astro}$,$N_{\rm Att}$, $N_{\rm Instr}$,$N_{\rm par}$, ...) \;
		\texttt{\textbf{aprod1\_Kernel\_glob<<<}}gridDim,blockDim\texttt{\textbf{>>>}}  \; \nonl ($\mathbf{A}_\textbf{d,dev}$,$\vb*{x}_\textbf{dev}$,$\vb*{b}_\textbf{dev}$,$N[pid]$,$N_{\rm Astro}$,$N_{\rm Att}$, $N_{\rm Instr}$,$N_{\rm Glob}$,$N_{\rm par}$, ...) \; \label{algo_aprod_1:c:kernels_call_in_main_scope_end}
		\texttt{\textbf{Constraints kernels<<<}}...\texttt{\textbf{>>>}}(...)\;
		\texttt{\textbf{cudaDeviceSynchronize}}() \; \label{algo:aprod_1:c:cudaDeviceSynchronize}
		\texttt{MPI\_Allreduce}($\vb*{b}$) \; \label{algo:aprod_1:c:Reduce}
		... \;
		\} \;
		
	\end{multicols}
	\caption{\textit{aprod} 1 with OpenMP, OpenACC, and CUDA\label{algo:aprod_1}}
\end{algorithm*}


\begin{algorithm*}
	\tiny
	\begin{multicols}{3}
		\nonl \textbf{\small \textit{aprod} 2 with OpenMP}\;
		\SetNlSty{}{o.}{}
		int main(int argc, char **argv) \;
		\{ \;
		... \;
		\texttt{\textbf{\#pragma omp parallel private($pid$,$tid$,$nth$) shared($\vb*{N}$,$\vb*{x}$,$\mathbf{A}_\textbf{d}$,$\vb*{b}$)}} \;
		\{\;
		\tcp{ID number of the OpenMP thread} 
		\text{ }\text{ }\text{ }\textbf{$tid$ = omp\_get\_thread\_num()}\; \label{algo:omp:aprod2_tid}
		\tcp{Number of OpenMP threads}
		\text{ }\text{ }\text{ }\textbf{$nth$ = omp\_get\_num\_threads()}\; \label{algo:omp:aprod2_nth}
		\text{ }\text{ }\text{ }\For{$i\leftarrow \vb*{N}_\textbf{t}[tid][0]$ \KwTo $\vb*{N}_\textbf{t}[tid][1]$}  { \label{algo:omp:aprod2_Ntid}
			\nonl \;
			\tcp{Astrometric sect.}
			$k = i \times N_{\rm par}$ \;
			\For{$j\leftarrow 0$ \KwTo $N_{\rm Astro}$}{
				\nonl \;
				$\vb*{x}[j+ {\rm offset}[i]] = \vb*{x}[j + {\rm offset}[i]] + \mathbf{A}_\textbf{d}[k]\vb*{b}[i]$\;
				$k$++\;
			}
			\nonl \;
			\tcp{Attitude sect.}
			$k = i \times N_{\rm par} + N_{\rm Astro}$ \;
			\For{$j_1\leftarrow 0$ \KwTo $N_{\rm Axes}$}{
				$k_2 = j_1 \times N_{\rm DFA} + {\rm offset}[i] $\;
				\For{$j_2 \leftarrow 0$ \KwTo $N_{\rm ParAxis}$}{
					$\vb*{x}[j_2 + k_2] = \vb*{x}[j_2 + k_2] + \mathbf{A}_\textbf{d}[k]\vb*{b}[i]$\;
					$k$++\;
				}  
			}
			\nonl \;
			\tcp{Instrumental sect.}	        
			$k = i \times N_{\rm par} + N_{\rm Astro} + N_{\rm Att}$ \;
			\For{$j\leftarrow 0$ \KwTo $N_{\rm Instr}$}{
				$\vb*{x}[\mathcal{F}(i,j)$+${\rm offset}]$ = $\vb*{x}[\mathcal{F}(i,j)$+${\rm offset}]$+$\mathbf{A}_\textbf{d}[k]\vb*{b}[i]$\;
				$k$++\;
			}
			\nonl \;
			\tcp{Global sect.}
			$k = i \times N_{\rm par} + N_{\rm Astro} + N_{\rm Att} + N_{\rm Instr}$ \;
			\For{$j\leftarrow 0$ \KwTo $N_{\rm Glob}$}{
				$\vb*{x}[j+ {\rm offset}] = \vb*{x}[j + {\rm offset}] + \mathbf{A}_\textbf{d}[k]\vb*{b}[i]$\; 
				$k$++\;
			}
		}
		\}\;
		Constraints computation\;
		\texttt{MPI\_Allreduce($\vb*{x}$)} \; \label{algo:aprod_2:o:Reduce}
		... \;
		\} \;
		
		\nonl \;
		\nonl \;
		\nonl \;
		\nonl \;
		\nonl \;
		\nonl \;
		\nonl \;
		\nonl \;
		\nonl \;
		\nonl \;
		\nonl \;
		\nonl \;
		\nonl \;
		\nonl \;
		\nonl \;
		\nonl \;
		\nonl \;
		\nonl \;
		\nonl \;
		\nonl \;
		\nonl \;
		\nonl \;
		\nonl \;
		\nonl \;
		\nonl \;
		\nonl \;
		\nonl \;
		\nonl \;
		\nonl \;
		\nonl \;
		\nonl \;
		\nonl \;
		\nonl \;
		\nonl \;
		\nonl \;
		\nonl \;
		\nonl \;
		\nonl \;
		\nonl \;
		\nonl \;
		\nonl \;
		\nonl \;
		\nonl \;
		\nonl \;
		\nonl \;
		\nonl \;
		\nonl \;
		\nonl \;
		\nonl \;
		\nonl \;
		\nonl \;
		\nonl \;
		\nonl \;
		\nonl \;
		\nonl \;
		\nonl \;
		\nonl \;
		\nonl \;
		\nonl \;
		\nonl \;
		
		\setcounter{AlgoLine}{0}
		\SetNlSty{}{a.}{}
		\nonl \textbf{\small \textit{aprod} 2 with OpenACC}\;
		
		int main(int argc, char **argv) \;
		\{ \;
		... \;
		\texttt{\textbf{\#pragma acc parallel}} \; \label{algo:acc:aprod_2_parallel_start} 
		\{\;
		\nonl \;
		\nonl \;
		\text{ }\text{ }\text{ }\texttt{\textbf{\#pragma acc loop}} \; \label{algo:acc:aprod_2_ext_loop}
		\text{ }\text{ }\text{ }\For{$i\leftarrow 0$ \KwTo $\vb*{N}[pid]$}  {
			\nonl \;
			\tcp{Astrometric sect.}
			$k = i \times N_{\rm par}$ \;
			\For{$j\leftarrow 0$ \KwTo $N_{\rm Astro}$}{
				\texttt{\textbf{\#pragma acc atomic}}\; \label{algo:acc:atomic_Astro}
				$\vb*{x}[j+ {\rm offset}[i]] = \vb*{x}[j + {\rm offset}[i]] + \mathbf{A}_\textbf{d}[k+j]\vb*{b}[i]$\; \label{algo:aprod_2:a:astro_operation}
			}
			\nonl \;
			\tcp{Attitude sect.}
			$k = i \times N_{\rm par} + N_{\rm Astro}$ \;
			\For{$j_1\leftarrow 0$ \KwTo $N_{\rm Axes}$}{
				$k_1 = j_1 \times N_{\rm ParAxis} $\;
				$k_2 = j_1 \times N_{\rm DFA} + {\rm offset}[i] $\;
				\For{$j_2 \leftarrow 0$ \KwTo $N_{\rm ParAxis}$}{
					\texttt{\textbf{\#pragma acc atomic}}\; \label{algo:acc:atomic_Att}
					$\vb*{x}[j_2 + k_2] = \vb*{x}[j_2 + k_2] + \mathbf{A}_\textbf{d}[k + j_2 + k_1]\vb*{b}[i]$\; \label{algo:aprod_2:a:att_operation}
				}  
			}
			\nonl \;   	
			\tcp{Instrumental sect.}        
			$k = i \times N_{\rm par} + N_{\rm Astro} + N_{\rm Att}$ \;
			\For{$j\leftarrow 0$ \KwTo $N_{\rm Instr}$}{
				\texttt{\textbf{\#pragma acc atomic}}\; \label{algo:acc:atomic_Instr}
				$\vb*{x}[\mathcal{F}(i,j) + {\rm offset}] = \vb*{x}[\mathcal{F}(i,j) + {\rm offset}] + \mathbf{A}_\textbf{d}[k+j]\vb*{b}[i]$\; \label{algo:aprod_2:a:instr_operation}
			}
			\nonl \;
			\tcp{Global sect.}
			$k = i \times N_{\rm par} + N_{\rm Astro} + N_{\rm Att} + N_{\rm Instr}$ \;
			\For{$j\leftarrow 0$ \KwTo $N_{\rm Glob}$}{ \label{algo:aprod_2:a:global_for_loop}
				\texttt{\textbf{\#pragma acc atomic}}\; \label{algo:acc:atomic_Glob}
				$\vb*{x}[j+ {\rm offset}] = \vb*{x}[j + {\rm offset}] + \mathbf{A}_\textbf{d}[k+j]\vb*{b}[i]$\; \label{algo:acc:xpluseqAb_Glob} \label{algo:aprod_2:a:global_operation}
				\nonl \; 
			}
		}
		\} \; \label{algo:acc:aprod_2_parallel_end}
		\nonl \;
		Constraints computation\;
		\texttt{MPI\_Allreduce($\vb*{x}$)} \; \label{algo:aprod_2:a:Reduce}
		... \;
		\} \;
		
		\nonl \;
		\nonl \;
		\nonl \;
		\nonl \;
		\nonl \;
		\nonl \;
		\nonl \;
		\nonl \;
		\nonl \;
		\nonl \;
		\nonl \;
		\nonl \;
		\nonl \;
		\nonl \;
		\nonl \;
		\nonl \;
		\nonl \;
		\nonl \;
		\nonl \;
		\nonl \;
		\nonl \;
		\nonl \;
		\nonl \;
		\nonl \;
		\nonl \;
		\nonl \;
		\nonl \;
		\nonl \;
		\nonl \;
		\nonl \;
		\nonl \;
		\nonl \;
		\nonl \;
		\nonl \;
		\nonl \;
		\nonl \;
		\nonl \;
		\nonl \;
		\nonl \;
		\nonl \;
		\nonl \;
		\nonl \;
		\nonl \;
		\nonl \;
		\nonl \;
		\nonl \;
		\nonl \;
		\nonl \;
		\nonl \;
		\nonl \;
		\nonl \;
		\nonl \;
		\nonl \;
		\nonl \;
		\nonl \;
		\nonl \;
		\nonl \;
		\nonl \;
		
		\setcounter{AlgoLine}{0}
		\SetNlSty{}{c.}{}
		\nonl \textbf{\small \textit{aprod} 2 with CUDA}\;
		
		\tcp{Astrometric sect.}
		\textbf{\texttt{\_\_global\_\_ void} \texttt{aprod2\_Kernel\_astro}\; \nonl (double* $\mathbf{A}_\textbf{d,dev}$, double* $\vb*{x}_\textbf{dev}$,\; \nonl double* $\vb*{b}_\textbf{dev}$, long $n$, \; \nonl short $N_\textbf{Astro}$, long $N_\textbf{par}$, ...)} \; \label{algo:aprod_2:c:CUDA_kernels_definition_start}
		\{ \;
		
		\text{ } \textbf{$i$ = \texttt{blockIdx.x*blockDim.x+threadIdx.x}}\; \label{algo:aprod_2:c:grid_x_astro}
		\text{ } {\bf if} ($i < n$)  \; \label{algo:aprod_2:c:if_astro}
		\text{ } \{ \;
		\text{ }\text{ } $k = i \times N_{\rm par}$ \;
		\text{ }\text{ } \For{$j\leftarrow 0$ \KwTo $N_{\rm Astro}$}{
			\textbf{\texttt{atomicAdd}(\&$\vb*{x}[j+ {\rm offset}[i]], \mathbf{A}_\textbf{d}[k+j]\vb*{b}[i]$)} \; \label{algo:aprod_2:c:atomic_astro}  
		}
		\text{ } \} \;
		\} \;
		
		\tcp{Attitude sect., axis 0}
		\textbf{ \texttt{\_\_global\_\_ void} \texttt{aprod2\_Kernel\_attA0} \; \nonl (double* $\mathbf{A}_\textbf{d,dev}$, double* $\vb*{x}_\textbf{dev}$,\; \nonl double* $\vb*{b}_\textbf{dev}$, long $n$, \; \nonl short $N_\textbf{DFA}$, short $N_\textbf{ParAxis}$, long $N_\textbf{par}$, short $N_\textbf{Astro}$ ...)} \;
		\{ \;
		
		\text{ } $k_1 = 0 \times N_{\rm ParAxis} $\;
		
		\text{ } \textbf{$i$ = \texttt{blockIdx.x*blockDim.x+threadIdx.x}}\; \label{algo:aprod_2:c:grid_x_att_0}
		\text{ } {\bf if} ($i < n$)  \; \label{algo:aprod_2:c:if_att_0}
		\text{ } \{ \;
		\text{ }\text{ } $k = i \times N_{\rm par} + N_{\rm Astro}$ \;
		\text{ }\text{ } $k_2 = 0 \times N_{\rm DFA} + {\rm offset}[i] $\;
		\text{ }\text{ } \For{$j_2 \leftarrow 0$ \KwTo $N_{\rm ParAxis}$}{
			\textbf{\texttt{atomicAdd}(\&$\vb*{x}[j_2 + k_2], \mathbf{A}_\textbf{d}[k + j_2 + k_1]\vb*{b}[i]$)} \;  \label{algo:aprod_2:c:atomic_att_0}
		}
		\text{ } \} \;
		\} \;
		
		\tcp{Attitude sect., axis 1}
		\textbf{\texttt{\_\_global\_\_ void} \texttt{aprod2\_Kernel\_attA1}(...)} \;
		\{...\} \;
		
		\tcp{Attitude sect., axis 2}
		\textbf{\texttt{\_\_global\_\_ void} \texttt{aprod2\_Kernel\_attA2}(...)} \;
		\{...\} \;
		
		\tcp{Instrumental sect.}
		\textbf{\texttt{\_\_global\_\_ void} \texttt{aprod2\_Kernel\_instr} \; \nonl (double* $\mathbf{A}_\textbf{d,dev}$, double* $\vb*{x}_\textbf{dev}$,\; \nonl double* $\vb*{b}_\textbf{dev}$, long $n$, \; \nonl short $N_{\rm Astro}$, short $N_{\rm Att}$, short $N_{\rm Instr}$, long $N_{\rm par}$, ...)} \;
		\{ \;
		
		\text{ } \textbf{$i$ = \texttt{blockIdx.x*blockDim.x+threadIdx.x}}\; \label{algo:aprod_2:c:grid_x_instr}
		\text{ } {\bf if} ($i < n$)  \; \label{algo:aprod_2:c:if_instr}
		\text{ } \{ \;
		\text{ }\text{ } $k = i \times N_{\rm par} + N_{\rm Astro} + N_{\rm Att}$ \;
		\text{ }\text{ } \For{$j\leftarrow 0$ \KwTo $N_{\rm Instr}$}{
			\textbf{\texttt{atomicAdd}(\&$\vb*{x}[\mathcal{F}(i,j) + {\rm offset}], \mathbf{A}_\textbf{d}[k+j]\vb*{b}[i]$)} \; \label{algo:aprod_2:c:atomic_instr}
		}
		\text{ } \} \;
		\} \;
		
		\tcp{Global sect., part 1}
		\textbf{\texttt{\_\_global\_\_ void} \texttt{aprod2\_Kernel\_globSum1}(double* $\mathbf{A}_\textbf{d,dev}$, double* $\vb*{x}_\textbf{dev}$, double* $\vb*{b}_\textbf{dev}$, long $n$, short $N_{\rm Astro}$,  short $N_\textbf{Att}$, short $N_\textbf{Instr}$, long $N_\textbf{par}$, double* $\vb*{x}_\textbf{dev,sum}$, int $j$, ...)} \; \label{algo:aprod_2:c:glob_1}
		\{...\} \;
		
		\tcp{Global sect., part 2}
		\textbf{\texttt{\_\_global\_\_ void} \texttt{aprod2\_Kernel\_globSum2}(double* $\vb*{x}_\textbf{dev}$, long \texttt{arraySize}, short $N_{\rm Astro}$, double* \texttt{gArr}, int $j$, ...)} \; \label{algo:aprod_2:c:glob_2}
		\{...\} \;
		
		\textbf{\texttt{Constraints kernels}}\; \label{algo:aprod_2:c:CUDA_kernels_definition_end}
		
		... \;
		
		\texttt{static const int} blockSize = 1024 \; \label{algo_aprod_2:c:grid_def_aprod_2_glob_1}
		\texttt{static const int} gridSize = 24*16 \; \label{algo_aprod_2:c:grid_def_aprod_2_glob_2}
		
		\nonl \;
		
		int main(int argc, char **argv) \;
		\{ \;
		... \;
		
		TW = 1024 \;
		\texttt{dim3} gridDim (($N[pid]$ - 1)/TW + 1,1,1) \;
		\texttt{dim3} blockDim (TW,1,1) \;
		
		\texttt{\textbf{aprod2\_Kernel\_astro<<<}}gridDim,blockDim,0,stream1\texttt{\textbf{>>>}} \; \nonl ($\mathbf{A}_\textbf{d,dev}$,$\vb*{x}_\textbf{dev}$,$\vb*{b}_\textbf{dev}$,$N[pid]$,$N_{\rm Astro}$,$N_{\rm par}$, ...) \; \label{algo_aprod_2:c:kernels_call_in_main_scope_start}
		\texttt{\textbf{aprod2\_Kernel\_attA0<<<}}gridDim,blockDim,0,stream2\texttt{\textbf{>>>}} \; \nonl ($\mathbf{A}_\textbf{d,dev}$,$\vb*{x}_\textbf{dev}$,$\vb*{b}_\textbf{dev}$,$N[pid]$,$N_{\rm DFA}$,$N_{\rm ParAxis}$, $N_{\rm par}$,$N_{\rm Astro}$ ...) \;
		\texttt{\textbf{aprod2\_Kernel\_attA1<<<}}gridDim,blockDim,0,stream3\texttt{\textbf{>>>}} \;  \nonl ($\mathbf{A}_\textbf{d,dev}$,$\vb*{x}_\textbf{dev}$,$\vb*{b}_\textbf{dev}$,$N[pid]$,$N_{\rm DFA}$,$N_{\rm ParAxis}$, $N_{\rm par}$,$N_{\rm Astro}$ ...) \;
		\texttt{\textbf{aprod2\_Kernel\_attA2<<<}}gridDim,blockDim,0,stream4\texttt{\textbf{>>>}} \; \nonl ($\mathbf{A}_\textbf{d,dev}$,$\vb*{x}_\textbf{dev}$,$\vb*{b}_\textbf{dev}$,$N[pid]$,$N_{\rm DFA}$,$N_{\rm ParAxis}$, $N_{\rm par}$,$N_{\rm Astro}$ ...) \;
		\texttt{\textbf{aprod2\_Kernel\_instr<<<}}gridDim,blockDim,0,stream5\texttt{\textbf{>>>}} \; \nonl ($\mathbf{A}_\textbf{d,dev}$,$\vb*{x}_\textbf{dev}$,$\vb*{b}_\textbf{dev}$,$N[pid]$,$N_{\rm Astro}$,$N_{\rm Att}$, $N_{\rm Instr}$,$N_{\rm par}$, ...) \; \label{algo_aprod_2:c:kernels_call_in_main_scope_before_global_part_end}
		\For{$j \leftarrow 0$ \KwTo $N_{\rm Glob}$}{ \label{algo:aprod_2:c:global_for_loop}
			\texttt{\textbf{aprod2\_Kernel\_globSum1<<<}}gridSize, blockSize\texttt{\textbf{>>>}} ($\mathbf{A}_\textbf{d,dev}$,$\vb*{x}_\textbf{dev}$,$\vb*{b}_\textbf{dev}$,$n$,$N_{\rm Astro}$,$N_\textbf{Att}$, $N_\textbf{Instr}$,$N_\textbf{par}$,$\vb*{x}_\textbf{dev,sum}$,$j$, ...) \;
			\texttt{\textbf{aprod2\_Kernel\_globSum2<<<}}1, blockSize\texttt{\textbf{>>>}} ($\vb*{x}_\textbf{dev}$,\texttt{gridSize},$N_{\rm Astro}$,$\vb*{x}_\textbf{dev,sum}$,$j$, ...) \;
		}
	    \texttt{\textbf{Constraints kernels<<<}}...\texttt{\textbf{>>>}}(...)\; \label{algo_aprod_2:c:kernels_call_in_main_scope_end}
	    \textbf{\texttt{cudaDeviceSynchronize}()} \; \label{algo:aprod_2:c:cudaDeviceSynchronize}
		\texttt{MPI\_Allreduce}($\vb*{x}$) \; \label{algo:aprod_2:c:Reduce}
		... \;
		\} \;
		
	\end{multicols}
	\caption{\textit{aprod} 2 with OpenMP, OpenACC, and CUDA\label{algo:aprod_2}}
\end{algorithm*}

\section{Performance tests}
\label{sec:Gaia_AVU_GSR_code_performance_tests}

The MPI + OpenMP code has been in production since 2014 and it has run on all the Tier0 systems of CINECA. It is currently running on M100, which has 980 compute nodes having the following features:

\begin{enumerate}
	\item 2 sockets of 16 physical cores each, of the type IBM POWER9 AC922, with a processor speed of 3.1 GHz. Each physical core corresponds to 4 virtual cores, with a total of 128 (2 $\times$ 16 $\times$ 4) virtual cores per node;
	\item 4 GPUs of the type NVIDIA Volta V100, with a memory of 16 GB each, connected with Nvlink 2.0;
	\item 256 GB of RAM.
\end{enumerate}

As shown by different runs, such as the performance tests illustrated in~\cite{Cesare_2022}, the MPI + OpenMP code runs in its optimal configuration when parallelized on 16 MPI processes per node and 2 OpenMP threads per MPI process. For a typical run for the production occupying a memory of 340 GB, parallelized on 2 nodes on 16 MPI processes + 2 OpenMP threads per node, and with a number of observations and of stars equal to $N_{\rm obs} = 1.8 \times 10^9$ and $N_{\rm stars} = 8.4 \times 10^6$, respectively, we achieve a convergence after $\sim$141000 iterations with an iteration time of $\sim$$4.23$~s, which results in a total elapsed time of $t_{\rm e, OMP} \simeq 166$~hours, namely about one week. However, this elapsed time is obtained for a system having a number of observations $\sim$$2$ orders of magnitude smaller than the number of observations expected for the final Gaia dataset ($\sim$$10^{11}$). When we will have to deal with such a large dataset, that will occupy $\sim$10-100 TB of memory, the time-to-solution would become $\sim$30-300 times larger, which will result in a far from optimal production. To manage these data sizes, a properly accelerated code is needed. For this purpose, we compared the performance of the MPI + CUDA and of the MPI + OpenMP codes on M100 for a set of systems with increasing size, measuring the acceleration factor of the CUDA code over the OpenMP code to verify whether the CUDA code was worth to be put in production.

We ran the OpenMP and the CUDA applications for different input datasets provided by the Data Processing Center of Turin (DPCT), which is supervised by the Aerospace Logistics Technology Engineering Company (ALTEC) in collaboration with the Astrophysics Observatory of Turin (INAF-OATO). These inputs are real Gaia datasets and they are employed for the production of the OpenMP code. The datasets have different sizes, occupying a memory of 40, 100, 300, and 350 GB, and each of them only computes some sections of the complete model. The 40 GB and 300 GB systems solve the attitude and the instrumental parts, the 100 GB system solves the astrometric part, and the 350 GB system solves the astrometric, the attitude, and the instrumental parts. As anticipated in Section~\ref{sec:Gaia_AVU_GSR_code_CUDA_porting}, no system solves the global part.

\begin{figure*}[hbt!]
	\centering
	\includegraphics[width=0.4\textwidth]{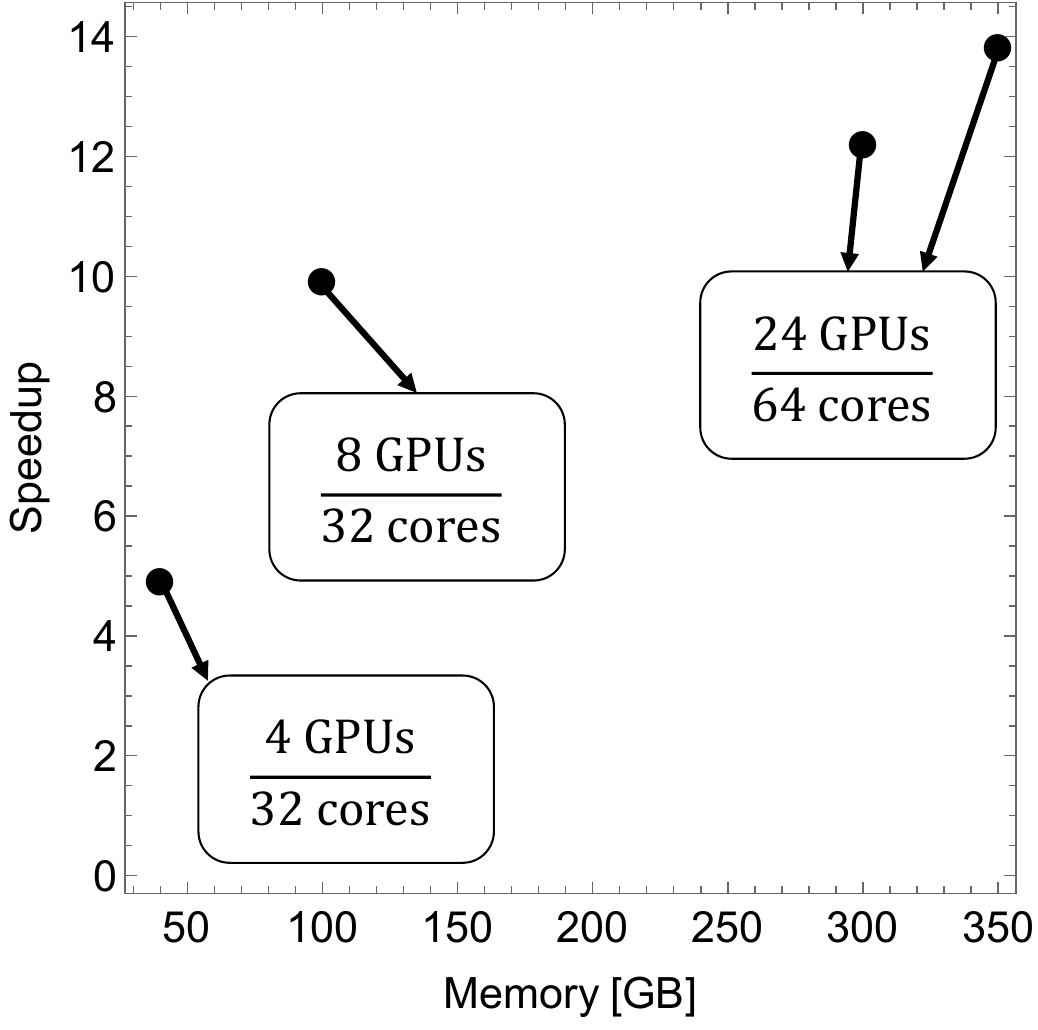}
	\caption{Speedup of the CUDA code over the OpenMP code as a function of the memory occupied by the system. For every point, the number of GPUs employed by the CUDA code and the number of physical cores employed by the OpenMP code is indicated.}
	\label{fig:Speedup_CUDA_over_OpenMP}
\end{figure*}

Figure~\ref{fig:Speedup_CUDA_over_OpenMP} shows the ratio between the average times of one LSQR iteration of the OpenMP and the CUDA codes, as a function of the system size. We ran the OpenMP and the CUDA codes in their optimal configurations (16 MPI processes + 2 OpenMP threads per node, for the OpenMP code, and 4 MPI processes per node for the CUDA code, see Section~\ref{sec:Gaia_AVU_GSR_code_CUDA_porting_Multi_GPU}). 
We report below each point in Figure~\ref{fig:Speedup_CUDA_over_OpenMP} the number of GPUs employed by the CUDA code, coincident with the number of MPI tasks, and of physical cores employed by the OpenMP code. Each code runs on the minimum number of nodes needed. For the same amount of memory, the CUDA code might require more nodes than the OpenMP code, since the memory of the four GPUs in each node is smaller than the RAM memory of the node (64 GB vs 256 GB). For example, the 100 GB system is paralleliized on 8 GPUs, i.e., 2 nodes, for the CUDA code, and on 32 cores, i.e., 1 node, for the OpenMP code.

The speedup of the CUDA code over the OpenMP code increases with both the system size and the number of employed GPU resources. From the first to the second point, correspondent to the 40 GB and the 100 GB systems, the speedup doubles, passing from $\sim$5 to $\sim$10. This might be explained by the fact that whereas the OpenMP code is always parallelized on the same amount of resources, in the CUDA code the number of resources doubles in the second run compared to the first run. Instead, considering the last two points, correspondent to the 300 GB and the 350 GB systems, the speedup does not substantially increase, passing from $\sim$12 to $\sim$14. In this case, the two systems are always parallelized on the same amount of resources. The slighly increase of the speedup might be justified by an increase of the GPU occupancy in the 350 GB system compared to the 300 GB system. In the 300 GB system, the memory assigned per MPI process, and thus per GPU, is of $\sim$9.5 GB, which implies a GPU occupancy of $\sim$60\%. Instead, in the 350 GB system, the memory assigned per MPI process is of $\sim$13 GB, which implies a GPU occupancy of $\sim$80\%.

\citet{Cesare_2022} show in Figures 4b and 5a a strong scaling test for the OpenMP code up to 16 nodes: the strong scaling curve already departs from the ideal linear speedup, tending to a plateau, after $\sim$3 nodes (96 physical cores). This means that, for a fixed amount of memory, the performance of the OpenMP code does not substantially improve if we continue to increase the number of physical cores on which it runs. The same figures show that the OpenACC strong scaling behavior is similar to the OpenMP one and that the ratio between the OpenMP and OpenACC mean iteration times is nearly constant and around 1.4. Given that in the CUDA code, as in the OpenACC code, the MPI tasks are assigned to the GPUs of the node in a round-robin fashion, we expect the strong scaling curve to be also similar for the CUDA code. Furthermore, the CUDA code is much more performant than the OpenACC code and, thus, we expect it to accelerate over the OpenMP code even if the latter is run on a larger amount of physical cores.

The maximum speedup of $\sim$14x is obtained for the 350 GB system. With this speedup, the CUDA code is more than 9x faster than the OpenACC code. Given the trend observed in Figure~\ref{fig:Speedup_CUDA_over_OpenMP}, we expect the speedup to continue to increase for systems of larger sizes. This is a remarkable result in perspective of the final dataset of Gaia which makes this implementation of the CUDA code a good candidate to be put in production. However, before proceeding, we performed a further test, detailed in the following section, to verify whether the rearrangement of the code required for the CUDA parallelization had impaired the correctness of the application.

\section{Numerical stability}
\label{sec:Gaia_AVU_GSR_code_Numerical_stability}

\begin{figure*}[hbt!]
	\centering
	\includegraphics[width=\textwidth]{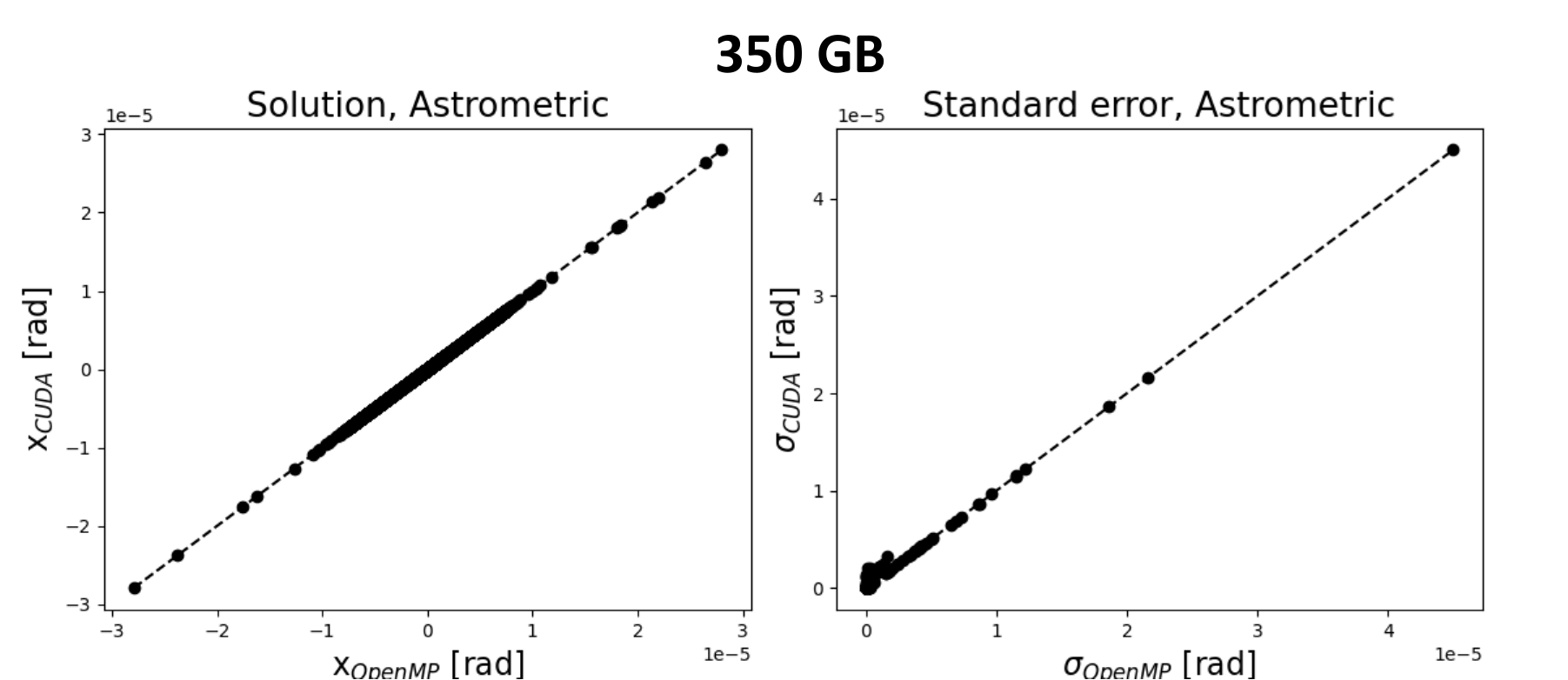}
	\caption{Solution (\textit{Figure~\ref{fig:Sol_StdErr_comp_350GB_Astro_LL}a}) of the astrometric section of the 350 GB system and its error (\textit{Figure~\ref{fig:Sol_StdErr_comp_350GB_Astro_LL}b}) computed with the CUDA code against the solution and the error of the same system computed with the OpenMP code. The one-to-one relation is shown as a black dashed line, for comparison.}
	\label{fig:Sol_StdErr_comp_350GB_Astro_LL}
\end{figure*}

To check if the CUDA parallelization was correctly implemented, we compared the solutions of the systems of equations considered in the previous section and the errors on the solutions, obtained with the OpenMP and the CUDA codes. Figure~\ref{fig:Sol_StdErr_comp_350GB_Astro_LL}a plots the solution of the astrometric section of the 350 GB system found with the CUDA code against the solution of the same system found with the OpenMP code. Figure~\ref{fig:Sol_StdErr_comp_350GB_Astro_LL}b shows the same for the errors on the solutions. The one-to-one relation (black dashed line) is shown as a reference. We do not illustrate the analogous plots for the other sections of the same system and for the other systems, since they show equivalent outputs.

The scatter plots in Figure~\ref{fig:Sol_StdErr_comp_350GB_Astro_LL} show that the CUDA and the OpenMP solutions and errors tightly distribute along the one-to-one relation, which suggests an agreement between the two couples of quantities. However, the figures only show a qualitative result, deduced by a visual inspection. To better quantify the consistency between the solutions and the errors found from the two codes, we calculate the average and the standard deviation of their differences, as reported in Table~\ref{tab:Mean_StDev_Diff_Solutions_StdErrors}. Table~\ref{tab:Mean_StDev_Diff_Solutions_StdErrors} also reports the same quantities for the attitude and instrumental sections of the 350 GB system and for the other systems. We can see that the average differences, both for the solutions ($d_x$) and for the errors ($d_\sigma$), are very close to 0, spanning a range, in absolute value, from $3.1 \times 10^{-23}$ rad \footnote{To simplify, from this point on we always write ``rad'' and ``arcsec'' and we do not distinguish between ``rad'' and ``rad~year$^{-1}$'' and between ``arcsec'' and ``arcsec~year$^{-1}$''.} to $1.5 \times 10^{-20}$ rad, for the solutions, and from $3.1 \times 10^{-14}$ rad to $3.2 \times 10^{-10}$ rad, for the standard errors. The standard deviations of the differences are, in every case, larger than the averages, which implies the agreement of the differences with zero. Moreover, the average differences, for both the solutions and the errors, are sometimes positive and sometimes negative, which suggests the absence of systematic errors. 

To better evaluate the agreement between the CUDA and the OpenMP solutions of every system, we also compared their differences with their errors. For each solution and error point, we computed the ratio:
\begin{equation}
\label{eq:sigma_check}
	q = \frac{\vert x_{i,{\rm CUDA}} - x_{i,{\rm OpenMP}} \vert}{\sqrt{\sigma^2_{i, {\rm CUDA}} + \sigma^2_{i, {\rm OpenMP}}}},
\end{equation}
where $x_{i,{\rm CUDA}}$ and $x_{i,{\rm OpenMP}}$ are the solution points found from the CUDA and the OpenMP codes, and $\sigma_{i, {\rm CUDA}}$ and $\sigma_{i, {\rm }OpenMP}$ are their errors. For every section of all the systems, the ratio $q$ is smaller than 1, which means that the solutions found from the two codes are always consistent within 1$\sigma$. These results show that the CUDA and the OpenMP solutions and errors are in agreement with each other for systems of increasing size and prove the numerical stability of the CUDA code. 

Besides checking the consistency between the results obtained with the two codes, we also wanted to verify if the solutions were obtained with the accuracy required by the Gaia mission ($\sim$$[10,100]$~$\mu$arcsec for the parallaxes and the positions and $\sim$$[10,100]$~$\mu$arcsec~year$^{-1}$ for the proper motions, see Section~\ref{sec:Intro}) to achieve a high precision astrometry, in order to properly investigate, e.g., the kinematics and the dynamics of the Galaxy. We converted the uncertainties on the solutions ($\sigma$) from radians to arcseconds with the relation:
\begin{equation}
\label{eq:rad2arcsec}
	\sigma ({\rm arcsec}) = \frac{\sigma ({\rm rad})}{4.84814 \times 10^{-6}}.
\end{equation}
In the 100 GB and 350 GB runs, which compute the astrometric section of the system, the average uncertainties on the astrometric parameters along with their standard deviations are of $\sigma = (4.0 \times 10^{-5} \pm 5.5 \times 10^{-3})$~arcsec and $\sigma = (2.1 \times 10^{-5} \pm 2.1 \times 10^{-4})$~arcsec, both for the OpenMP and the CUDA codes, in agreement with the needed precision. 
For the 100 GB run, nearly 80\% of the astrometric solution points have uncertainties below 10~$\mu$arcsec, and more than 97\% and 99\% of the astrometric solution points have uncertainties below 100~$\mu$arcsec and 500~$\mu$arcsec. For the astrometric part of the 350 GB run, the $\sim$99\% of the solution points already have uncertainties below 100~$\mu$arcsec. Also the attitude and instrumental parameters are generally obtained with a compatible accuracy.

Given these results, we put the CUDA code in production in Q2 2022. The CUDA solver was also put on a proprietary GitLab repository of CINECA and its copyright is held by INAF.

\begin{table*}[h!] 
	\begin{center}
		\begin{threeparttable}[b]
			\caption{Comparison between the CUDA and OpenMP solutions and errors on the solutions for the four considered systems of equations.}
			\label{tab:Mean_StDev_Diff_Solutions_StdErrors}
			\begin{tabular}{llcc}
				\toprule
				\toprule
				Memory & Section & $Mean(d_x) \pm StDev(d_x)$ & $Mean(d_\sigma) \pm StDev(d_\sigma)$  \\
				$[{\rm GB}]$ & & $[{\rm rad}]$ & $[{\rm rad}]$ \\
				(1) & (2) & (3) & (4) \\
				\midrule
				40 & Attitude & $3.1 \times 10^{-20} \pm 7.4 \times 10^{-17}$ & $-3.2 \times 10^{-10} \pm 1.5 \times 10^{-8}$ \\
				     & Instrumental & $7.3 \times 10^{-23} \pm 4.1 \times 10^{-20}$ & $-4.5 \times 10^{-11} \pm 1.5 \times 10^{-10}$ \\
				\midrule
				100 & Astrometric & $-1.5 \times 10^{-20} \pm 4.7 \times 10^{-17}$ & $3.1 \times 10^{-14} \pm 2.2 \times 10^{-12}$ \\
				\midrule
				300 & Attitude & $1.7 \times 10^{-21} \pm 3.7 \times 10^{-17}$ & $6.9 \times 10^{-11} \pm 1.2 \times 10^{-8}$ \\
				       & Instrumental & $-7.6 \times 10^{-23} \pm 1.5 \times 10^{-20}$ & $-1.9 \times 10^{-12} \pm 6.4 \times 10^{-12}$ \\
				\midrule
				350 & Astrometric & $-3.9 \times 10^{-22} \pm 2.2 \times 10^{-17}$ & $-4.0 \times 10^{-13} \pm 8.1 \times 10^{-10}$ \\
			           & Attitude & $1.2 \times 10^{-21} \pm 1.4 \times 10^{-21}$ & $-7.2 \times 10^{-13} \pm 5.7 \times 10^{-11}$ \\
				       & Instrumental & $3.1 \times 10^{-23} \pm 3.4 \times 10^{-20}$ & $1.2 \times 10^{-13} \pm 7.4 \times 10^{-13}$ \\
				\bottomrule
			\end{tabular}
			\begin{tablenotes}
				\item Column 1: Memory occupied by the system of equations; column 2: section solved for the considered system; column 3: mean and standard deviation of the differences between the solutions of the systems of equations found from the CUDA and the OpenMP codes; column 4: mean and standard deviation of the differences between the errors on the solutions found from the CUDA and the OpenMP codes.
				\item The quantities $d_x$ and $d_\sigma$ refer to the differences between a CUDA and an OpenMP quantity.
			\end{tablenotes}
		\end{threeparttable}
	\end{center}
\end{table*}

\section{Conclusions and future works}
\label{sec:Conclusions_and_future_works}

We ported to a GPU environment with the CUDA programming language the AVU--GSR parallel solver, developed for the ESA Gaia mission and originally parallelized on the CPU with a hybrid MPI + OpenMP model. The code solves a system of linear equations with the iterative LSQR algorithm to find the astrometric parameters of $\sim$$10^8$ stars in the Milky Way, the attitude and the instrumental settings of the Gaia spacecraft, and the global parameter $\gamma$ of the PPN formalism. To iteratively find the solution up to the convergence of the algorithm defined in the least square sense, the LSQR calls, at each step, the \textit{aprod} function in its modes 1 and 2, which provide an iterative estimate for the known terms and the solution arrays, respectively.

The porting presented in this paper is the result of an optimization of a previous GPU porting of this application, performed with the high-level language OpenACC. The OpenACC code showed a moderate speedup of $\sim$1.5x over the OpenMP code. As already pointed out at the beginning of Section~\ref{sec:Gaia_AVU_GSR_code_CUDA_porting}, further speedups might as well have been obtained with a better optimization of the usage of the OpenACC language. Indeed, the speedup of $\sim$1.5x refers to a quite basic parallelization with OpenACC, where the OpenMP directives were basically replaced by the OpenACC correspondent ones. However, we preferred to adopt the low-level language CUDA for the new porting since it allows to better match the architecture of the device and, thus, to possibly achieve larger performances. On the other hand, this reduces the code portability, since the CUDA parallelization is architecture-dependent. However, since the Gaia mission is expected to end in the following years and only a further porting of this code on Leonardo supercomputer is expected, we aimed to improve the performance rather than to obtain a larger code portability.

With the CUDA porting, we reorganized the structure of the Gaia AVU--GSR solver, by defining the kernels to parallelize different regions of the code, such as the \textit{aprod 1} and \textit{2} functions. In each of the kernels, we manually defined the hierarchy of the grid of threads to match as better as possible the GPU architecture and the topology of the problem to solve. We also ported with CUDA other regions of code that in the OpenACC application were still running on the CPU and we reduced the H2D and D2H data copies with respect to the OpenACC code. With these optimizations, the time fraction of one LSQR iteration due to GPU computation rises from $\sim$70\% to $\sim$90\%, and the time fractions due to CPU calculations and data transfers reduces from $\sim$15\% to $\sim$3\%.

Running the CUDA and the OpenMP applications on M100, the CUDA code presents a speedup over the OpenMP code increasing with the system size and with the employed GPU resources. The speedup reaches a maximum of $\sim$14 for a system occupying 350 GB of memory and is expected to increase for systems of larger sizes and by running the codes on next-generation platforms with GPUs having more memory and streaming multiprocessors, such as the CINECA supercomputer Leonardo. Indeed, the A200 GPUs of Leonardo have 4x more memory and more streaming multiprocessors than the V100 GPUs of M100, which allows to execute more concurrent threads. Since both M100 and Leonardo have 4 GPUs per node, the GPU memory per node on Leonardo is quadrupled with respect to M100. We plan to perform the first tests of the AVU--GSR code on Leonardo in the first half of 2023. 

The CUDA code showed great numerical stability, since it provided solutions and uncertainties on the solutions fully consistent within 1$\sigma$ with the correspondent ones found with the OpenMP code for a set of systems. Moreover, the solutions are obtained with the accuracy of $[10,100]$ $\mu$arcsec, as required by the Gaia mission. Given these results, the MPI + CUDA AVU--GSR solver was put in production on M100. This is a fundamental achievement since it provides an optimal production for the AVU--GSR pipeline, allowing to obtain important data for scientific analyses, such as the study of the Milky Way formation and evolution, in reduced timescales. 

The increasing trend of the speedup with the system size is a very important result toward the scientific purposes of the upcoming Data Releases of the Gaia mission, from which TBs of data will be produced up to an expected final dataset of $\sim$10-100 TB. In perspective of these pre-Exascale data products, we will continue the optimization process of the AVU--GSR code, for example by porting to the GPU further sections of code, and the consequent investigation of the performance, scaling, and numerical stability of the code, for systems with an increasing size, up to the sizes expected for the final Gaia dataset. These are some of the targets of a two-years project already underway and funded by INAF, an INAF Mini Grant, of which the author VC is the PI and which is performed in collaboration with Prof. Marco Aldinucci of the University of Turin. For this future analysis, we will use Leonardo, to better investigate the behavior of the AVU--GSR code on a next-generation pre-Exascale infrastructure and in perspective of a final porting of this code on Leonardo. 

Besides allowing a better performance, this novel arrangement of the hardware, with hosts less performant than the accelerators and GPUs with a larger memory and number of streaming multiprocessors, such as on Leonardo, will imply a low energy consumption for the size of the problems that will need to be computed by HPC GPU-oriented applications. 
When a code such as the AVU--GSR solver is ported to the GPU resulting in a $\gtrsim 14$x speedup over the CPU version, besides obtaining results in a minor time, we also expect to save a substantial amount of energy. This is not obvious, since H2D and D2H data transfers, not present in the CPU application, might be rather energy consuming, but this might be compensated by the high speedup. In a future work, we aim to compare the energy consumption of the CUDA and the OpenMP codes by running systems of increasing size, up to $\sim$10-100 TB, both on M100 and on Leonardo, to verify whether the CUDA code is in fact ``greener'' than the OpenMP code and whether running on Leonardo allows to save even more energy than on M100. Since the GPU memory per node on Leonardo is 4x the GPU memory per node on M100, a quarter of the resources could be required on Leonardo with respect to M100 to run a system of equal size. This might imply a minor energy consumption on Leonardo compared to M100. Also this analysis is a target of the Mini Grant project.

This research activity has important repercussions in the developement, toward a (pre-)Exascale calculation, of other LSQR-based applications involving the solutions of systems with a high sparsity degree, similarly to the Gaia AVU--GSR solver. The parallelization techniques employed in this code could be adapted and exploited in different contexts that adopt the LSQR, such as the reconstruction of images in radioastronomy~\citep{Naghibzadeh_and_vanderVeen_2017}, geophysics~\citep{Joulidehsar_2018,LIANG_2019,LSQR_geology_2019}, geodesy~\citep{Baur_and_Austen_2005}, medicine~\citep{Bin_2020,Guo_2021}, and industry~\citep{Jaffri_2020} (see Section~\ref{sec:Intro}). In conclusion, the continue developement of efficient parallelization techiques is essential to face the increasingly faster production of data in contexts of different nature, going toward the Big Data era.

\section*{Acknowledgements}
\label{sec:Acknowledgements}

    We sincerely thank the referee, whose comments largely improved and clarified the presentation of our results.
    We sincerely thank Dr. Aswin Kumar of NVIDIA for his support in parallelizing this application with CUDA, and the organizers of the CINECA course ``Programming paradigms for GPU devices'', held on 2021 June 9th-11th, for providing the material employed to learn the most important notions necessary to parallelize this code with CUDA. We also thank Dr. Massimiliano Guarrasi of CINECA, for deepening some CUDA conceps.
    This work has been supported by the Spoke 1 ``FutureHPC \& BigData'' of the ICSC--Centro Nazionale di Ricerca in High Performance Computing, Big Data and Quantum Computing-and hosting entity, funded by European Union--Next GenerationEU.
    This work was also supported by the Italian Space Agency (ASI) [grant No.: 2018-24-HH.0], in support of the Italian participation to the Gaia mission, and by Consorzio Interuniversitario Nazionale per l'Informatica, under the project EUPEX, EC H2020 RIA, EuroHPC-02-2020 [Grant Agreement: 101033975].

\bibliography{bib_Gaia_AVU_GSR_Parallel_CUDA}

\end{document}